\documentclass[twocolumn,secnumarabic,amssymb, nobibnotes, aps, prd,natbib]{revtex4-1}
\usepackage{graphicx}
\usepackage{subfigure}
\usepackage{amsmath}
\usepackage{hyperref}
\usepackage{aasmacros}

\usepackage{color}

\newcommand{\Pkdgrav}{\textsc{PkdGrav3}}
\begin{document}

\title{Cosmological constraints with deep learning from KiDS-450 weak lensing maps}

\author{Janis Fluri$^\mathrm{a}$}%
\email{janis.fluri@phys.ethz.ch}
\author{Tomasz Kacprzak$^\mathrm{a}$}
\author{Aurelien Lucchi$^\mathrm{b}$}
\author{Alexandre Refregier$^\mathrm{a}$}
\author{Adam Amara$^\mathrm{a}$}
\author{Thomas Hofmann$^\mathrm{b}$}
\author{Aurel Schneider$^\mathrm{a}$}
\affiliation{\vspace*{10pt}$^\mathrm{a}$Institute of Particle Physics and Astrophysics, Department of Physics, ETH Zurich, Switzerland}
\affiliation{$^\mathrm{b}$Data Analytics Lab, Department of Computer Science, ETH Zurich, Switzerland}
\date{\today}%

\defcitealias{Hildebrandt2016}{HV16}

\begin{abstract}

Convolutional Neural Networks (CNN) have recently been demonstrated on synthetic data to improve upon the precision of cosmological inference. In particular they have the potential to yield more precise cosmological constraints from weak lensing mass maps than the two-point functions.
We present the cosmological results with a CNN from the KiDS-450 tomographic weak lensing dataset, constraining the total matter density $\Omega_\mathrm{m}$, the fluctuation amplitude $\sigma_8$, and the intrinsic alignment amplitude $A_{\rm{IA}}$.
We use a grid of N-body simulations to generate a training set of tomographic weak lensing maps.
We test the robustness of the expected constraints to various effects, such as baryonic feedback, simulation accuracy, different value of $H_0$, or the lightcone projection technique.

We train a set of ResNet-based CNNs with varying depths to analyze sets of tomographic KiDS mass maps divided into 20 flat regions, with applied Gaussian smoothing of $\sigma=2.34$ arcmin.
The uncertainties on shear calibration and $n(z)$ error are marginalized in the likelihood pipeline.
Following a blinding scheme, we derive constraints on $S_8 = \sigma_8 (\Omega_\mathrm{m}/0.3)^{0.5} = 0.777^{+0.038}_{-0.036}$ with our CNN analysis, with $A_{\rm{IA}}=1.398^{+0.779}_{-0.724}$.
We compare this result to the power spectrum analysis on the same maps and likelihood pipeline and find an improvement of about $30\%$ for the CNN.
We discuss how our results offer excellent prospects for the use of deep learning in future cosmological data analysis.
\end{abstract}

\maketitle

\section{Introduction}

The current cosmological model used to describe the evolution of our universe suggests that we live in a spatially flat $\Lambda$CDM cosmology. In this model, small initial fluctuations in the matter distribution collapsed through gravitational interactions and formed highly non-linear structures that can be observed today. In weak gravitational lensing (WL) (see e.g. \cite{Schneider2005,Kilbinger2015review} for reviews) we aim to probe such structures directly through their interaction with the light from faint background galaxies. According to Einstein's theory of general relativity, matter interacts with the space-time and deflects the incoming photons, thereby slightly distorting the original source image. WL surveys such as the Canada France Hawaii Telescope Lensing Survey (CFHTLenS)\footnote{\url{cfhtlens.org}} \cite{CFHTLenS2013}, the Kilo-Degree Survey (KiDS)\footnote{\url{kids.strw.leidenuniv.nl}} \cite{Hildebrandt2018viking}, the Dark Energy Survey (DES)\footnote{\url{darkenergysurvey.org}} \cite{DESY1res}, and Subaru Hyper Suprime-Cam (HSC)\footnote{\url{hsc.mtk.nao.ac.jp/ssp/survey}} \cite{Hikage2019subaru} have already successfully applied this method to constrain cosmological parameters. Future surveys like Euclid \cite{Euclid2011} or LSST \cite{Chang2013} will be able to provide even more precise measurements.

Currently, the most common way to analyze the abundance of available data is the two-point correlation function (eg. \cite{Hildebrandt2018viking, DESY1res}). The two-point correlation function is able to perfectly describe the statistics of Gaussian random fields.
However, it lacks the capability of extracting non-Gaussian information that can significantly improve the constraints on the cosmological parameters, which lead to a search for new summary statistics.
Approaches based on weak lensing peak statistics (e.g. \cite{Dietrich2010peaks,Kacprzakpeaks, Fluri2018peak, liu:hal-01439964, KiDspeaks1, KiDspeaks2}) or the three-point correlations function (e.g. \cite{threepoint1, threepoint2}) have been applied to obtain cosmological constraints.
Recently, deep learning techniques have gained a lot of attention, due to their ability to automatically extract relevant features from image data, while being robust to noise.

A first demonstration of the ability of convolutional neural networks (CNN) to discriminate between different cosmologies directly from noisy weak lensing maps was done in \cite{Schmelzle2017}. Afterwards, \cite{Gupta2018} showed that one can use CNNs to predict the total matter density $\Omega_\mathrm{m}$ and the fluctuation amplitude $\sigma_8$ from a wide grid of different cosmologies using a parameter regression approach. In our previous work \cite{Fluri2018}, we examined the performance of a CNN compared to the traditional power spectrum analysis in the case of noisy convergence maps. And recently \cite{Ribli2019} performed a similar analysis using noisy convergence maps where they bench-marked different architectures of CNNs and compared their results to a power spectrum and weak lensing peak analysis.

Deep learning has recently been applied to various other astrophysical problems, such as fast finding of strongly lensed systems \citep{Lanusse2018deeplens}, measuring parameters of early galaxies using the 21-cm signal \citep{Gillet2019cm21}, fast Point Spread Function modeling \cite{Herbel2018}, the introduction of Baryonic effects into dark matter N-body simulations \cite{baryonpainting}, and other problems in cosmology \citep{Merten2019dissection,Ciuca2019string,Peel2018modifed,Gheller2018diffuse,LucieSmiuth2018structure,Rodriguez2018gan}.
And recent work on information maximizing networks \cite{IMNets}, neural density estimators, and the usage of networks to find optimal data compression \cite{Alsing_1, Alsing_2, Alsing_3} shows the potential of ML to improve the inference of cosmological parameters.

In this work, we perform a tomographic analysis of the KiDS-450 \citep[][hereafter \citetalias{Hildebrandt2016}]{Hildebrandt2016} data using a CNN to constrain  $\Omega_\mathrm{m}$, $\sigma_8$, and the intrinsic alignment amplitude $A_{\rm{IA}}$.
To our knowledge, this is the first time that a CNN was used to measure the cosmological parameters from observed weak lensing data.
We train our CNNs on a large number of tomographic lensing mass maps that include realistic masks and noise maps.
We apply a careful treatment of multiple systematic effects that can arise through astrophysical effects, measurement process, or during the construction of the simulations.
We marginalize the photometric redshift error, the multiplicative and additive shear biases, and the intrinsic alignment amplitude.
Further, we test if our inference pipeline is robust to other systematic effects such as the simulation algorithm, shear projection technique, or baryonic feedback \cite{Osato2015, Schneider2018}.
Following a blinding scheme, we derive constraints with a CNN in a number of configurations.
We compare the constraints from the CNN with those we obtain with a power spectrum analysis on the same maps, as well as with previous measurements on the KiDS-450 dataset.

The paper is structured in the following way. In section \ref{sec:data} we present an overview of the data used in the analysis. In section \ref{sec:blinding} we explain our blinding scheme. The methodology is described in section \ref{sec:methods}. This section contains the detailed description of the used N-body simulations (section \ref{sec:sims}) and all our considered systematic effects (section \ref{sec:systematics}). The examined networks are presented in section \ref{subsec:network} and the inference procedure is described in section \ref{sec:inference}. This is followed by the examination of the impact of the systematic effects in section \ref{sec:effects_sys} and the results of the cosmological analysis in section \ref{sec:results}. Appendix \ref{sec:multi_vs_single_cnn} describes how we combined the out of different networks. In appendices \ref{ap:loss} and \ref{ap:evals} we give further details regarding the training of the networks. Appendix \ref{ap:IA_amp} gives more insight about the constrained intrinsic alignment amplitude and appendix \ref{ap:specs} shows additional plots of the power spectrum analysis.

\section{KiDS-450 data \label{sec:data}}

In this work, we analyze the first 450 deg$^2$ of the Kilo Degree Survey (KiDS). KiDS is an ESO public survey that will, after completion, cover 1350 deg$^2$ in four bands (\emph{ugri}). It uses the OmegaCAM CCD mosaic camera mounted at the Cassegrain focus of the VLT Survey Telescope (VST). This combination of telescope and camera was specifically designed for weak lensing measurements, having a well-behaved and almost round point spread function and a small camera shear. The shear of the observed objects was estimated using \textsc{Lensfit} \cite{Miller2013} and a detailed description of the procedure can be found in \cite{Conti2017}. \textsc{Lensfit} uses a model of the point spread function at pixel level of individual exposures and extracts the ellipticities of galaxies by fitting a disc and bulge model via a likelihood-based method. This method enables it to provide weights for each object based on the likelihood of the fit. An estimate of the redshift of each object is obtained using the Bayesian photometric code \textsc{BPZ} from \cite{Benitez2000,Hildebrandt2012photoz}. Further, \citetalias{Hildebrandt2016} examined three different calibration methods to obtain the effective redshift distribution of the observed objects.

We use the publicly available KiDS-450 data that contains the shapes of roughly $\sim$15 million galaxies and is divided into five patches \cite{Jong2017}. The whole catalog has an effective galaxy number density of 8.53 galaxies/arcmin$^2$ and an ellipticity dispersion of $\sigma_e = 0.29$ per shear component.

\subsection{Data Preparation}

We split the galaxies into the same four tomographic bins using the \textsc{BPZ} estimate of their redshift, in the same way as \citetalias{Hildebrandt2016}.
\begin{figure}
\includegraphics[width=0.5\textwidth]{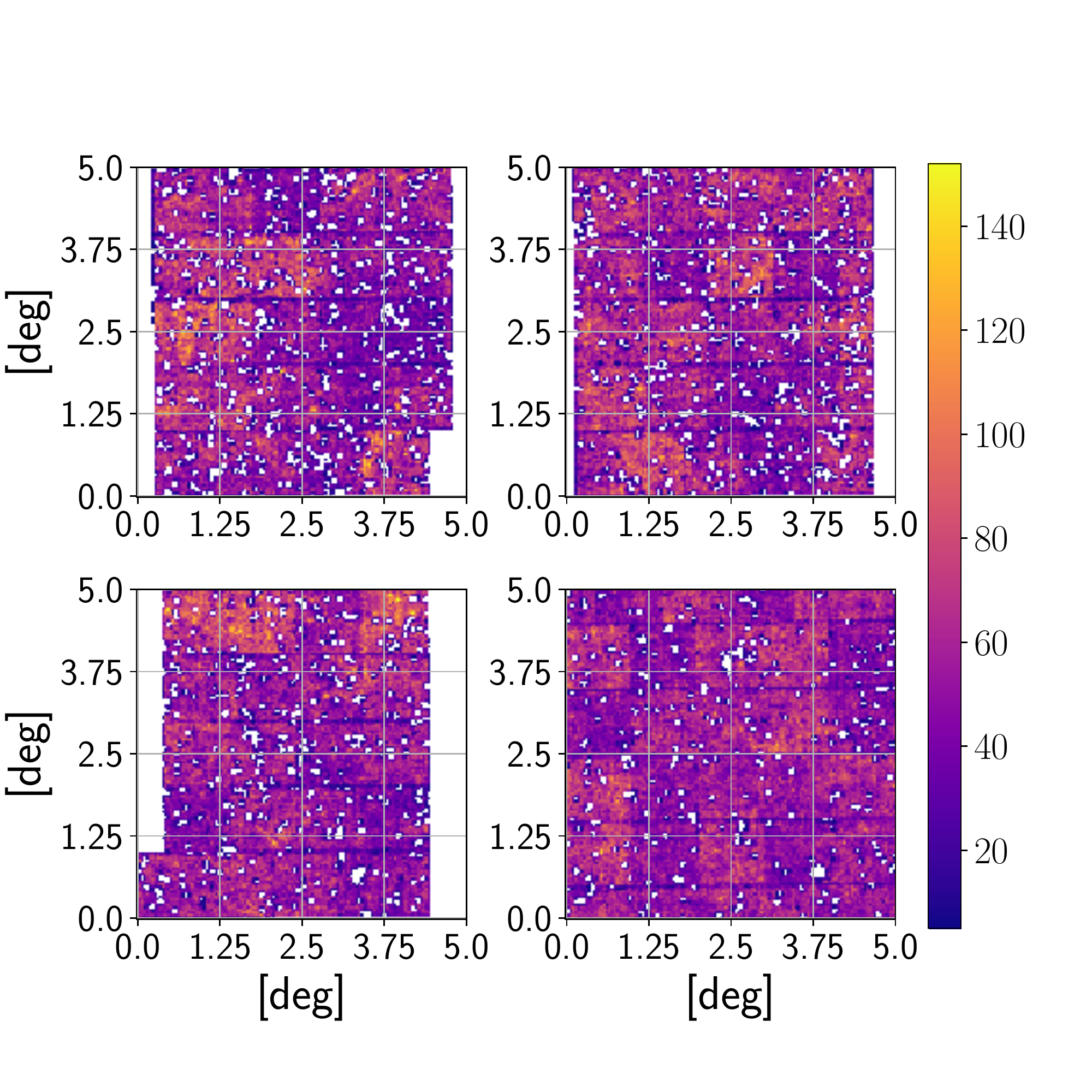}
\caption{The number of objects per pixel summed over all four redshift bins for four of our 20 projected 5$\times$5 deg$^2$ patches. \label{fig:patches}}
\end{figure}
\begin{table}
\centering
\begin{tabular}{|c|c|c|}
\hline
$z$ cuts & Number of Objects & After Projection  \\
\hline
$0.1 < z \leq 0.3$ & 3'879'822 & 3'268'559 \\
$0.3 < z \leq 0.5$ & 2'990'095 & 2'522'507 \\
$0.5 < z \leq 0.7$ & 2'970'569 & 2'501'016 \\
$0.7 < z \leq 0.9$ & 2'687'130 & 2'269'380 \\
\hline
Total & 12'527'616 & 10'561'462 \\
\hline
\end{tabular}
\caption{Number of object in the KiDS-450 data-set per redshift bin and the number of object used in this analysis to project the 20 different patches. \label{tab:zbins}}
\end{table}
The redshift bins have a width of $\Delta z = 0.2$ and only galaxies with a \textsc{BPZ} redshift estimate between $z = 0.1$ and $z = 0.9$ were used. Detailed information about the redshift bins is listed in table \ref{tab:zbins}. After these redshift cuts approximately $12.5$ million objects remained in the catalog.
We decided to adopt a similar approach as in our previous work \citep{Fluri2018} and projected the data onto smaller patches, to analyze the data with a CNN.
One should note that there already exist a suitable neural network architecture working on spherical data \cite{Perraudin2018}, which does not rely on smaller patches or the small angle approximation.
While this approach may prove necessary for large area surveys, like DES, LSST or Euclid, the KiDS-450 area can still be conveniently analyzed by splitting the area into relatively small number of patches. This approach also reduces the computational costs of the necessary simulations to train the network.
We split the data into 20 patches of $5\times 5$ deg$^2$ with a resolution of 128 pixels ($\sim2.34$ arcmin) per side. Some of those patches are shown in figure \ref{fig:patches}. We used the gnomonic projection and the estimated shear value of each pixel was calculated with the \textsc{Lensfit} weights $w_i$ and the measured ellipticity $e_i$ of each galaxy
\begin{equation}
\gamma_\mathrm{pix} = \frac{\sum_{i \in \mathrm{pix}}w_i (e_i - c_z)}{\sum_{i \in \mathrm{pix}}w_i}, \label{eq:shear}
\end{equation}
where $c_z$ is the additive shear bias (section \ref{subsec:shearbias}) of the corresponding redshift bin and projected patch. Due to the sparse distribution of the objects, it was not possible to fit the complete catalog inside the 20 patches and only $\sim 10.5$ million objects were used to create the shear maps.
The generated shear maps were then fed into the CNN to predict the underlying cosmological parameters.

\section{Blinding}
\label{sec:blinding}

The data described in the previous sections was used to generate noise maps (see section \ref{subsec:noise}), observation masks and the estimates of the additive and multiplicative shear biases (see section \ref{subsec:shearbias}).
However, to avoid confirmation bias, we calculated the cosmology constraints using the shear signal only after all the tests of the inference pipeline on simulations were completed and the design of the analysis was finalized (see section \ref{sec:methods}) and did not alter our cosmological analysis after this.

\section{Methodology \label{sec:methods}}

The CNN-based analysis required a large and accurate set of simulations that include noise and various systematic effects.
\begin{figure}
\includegraphics[width=0.3\textwidth]{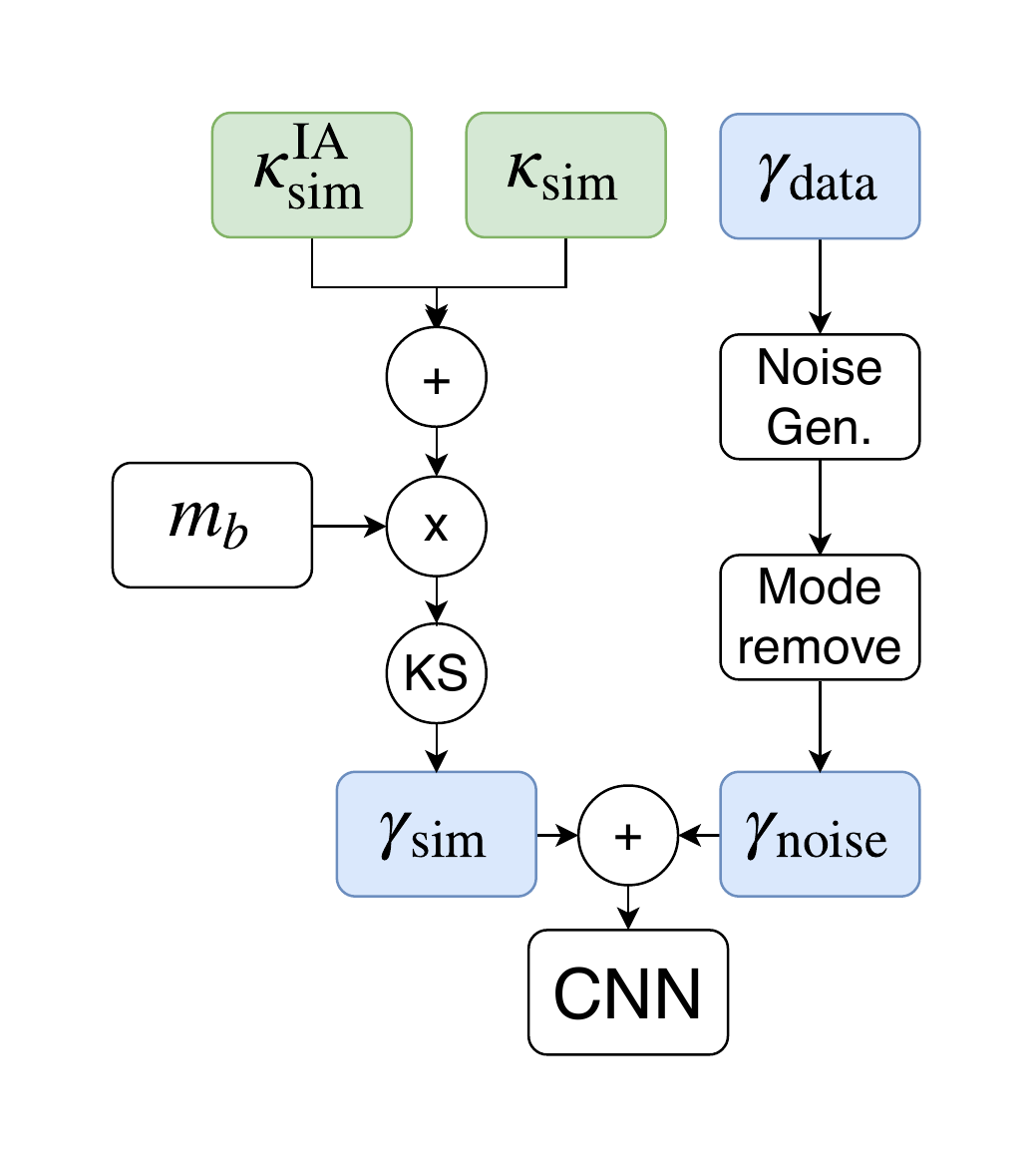}
\caption{Overview of the training pipeline.
The different blocks are described in section \ref{sec:methods}.
\label{fig:pipeline}}
\end{figure}
A schematic diagram of the training pipeline is shown in figure \ref{fig:pipeline}. The simulated convergence maps (section \ref{sec:sims}) and intrinsic alignment maps (section \ref{subsec:IA}) are combined with a random intrinsic alignment amplitude drawn from our prior. Furthermore, we model the multiplicative bias (section \ref{subsec:shearbias}) and transform the resulting convergence maps into shear maps using the Kaiser Squires (KS) \cite{KS1993} inversion and mask them (section \ref{subsec:KSinversion}). The observed data is used to generate random noise realizations (section \ref{subsec:noise}) that, after we apply a correction explained in section \ref{subsec:KSinversion}, are added to the simulated shear maps. These noisy shear maps are then used to train the CNN (section \ref{subsec:network}).

\subsection{Simulations \label{sec:sims}}

\subsubsection{N-body Simulations}

\begin{figure}
\includegraphics[width=0.5\textwidth]{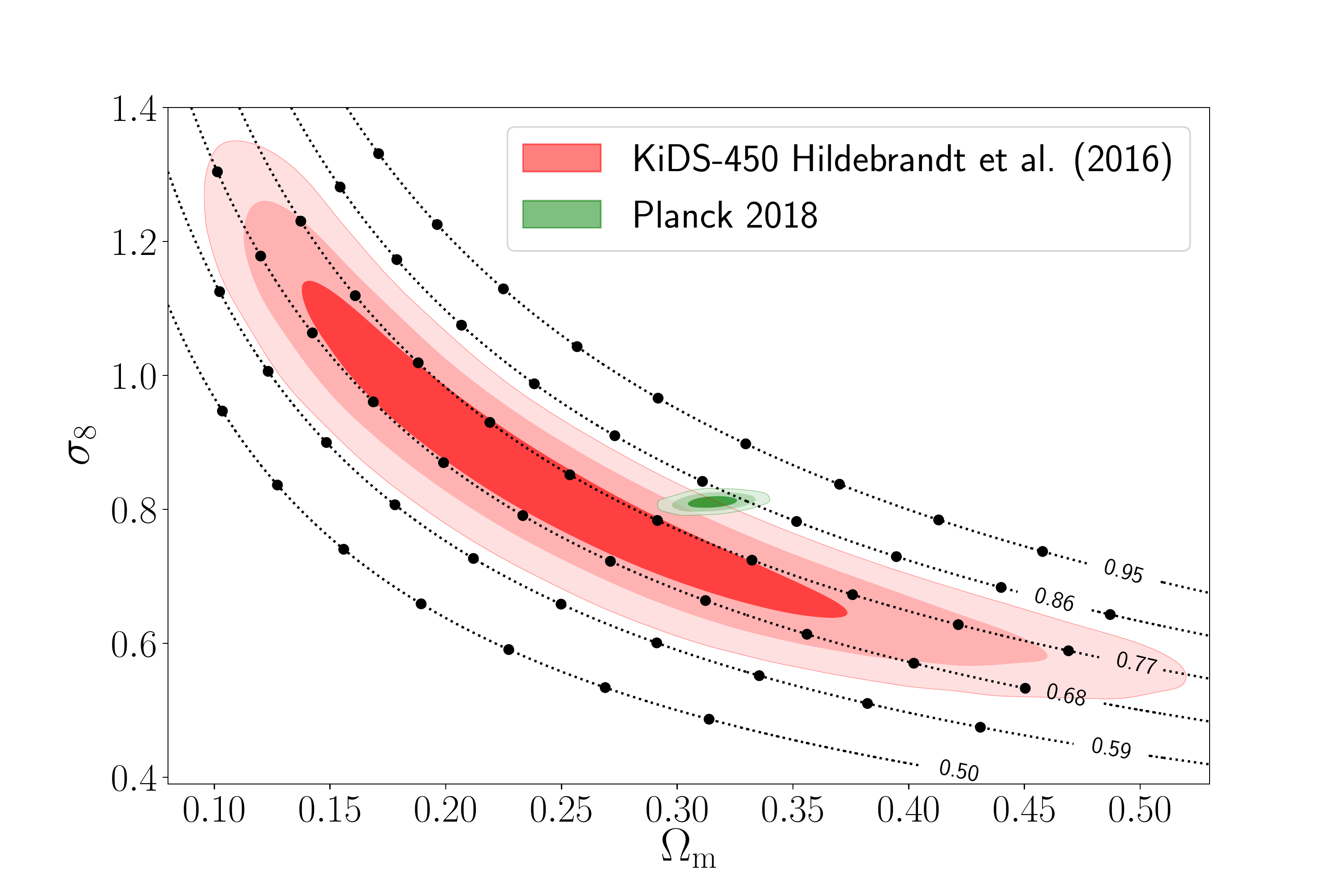}
\caption{The grid of simulated cosmologies. The dotted lines show the sampled values of $\Sigma_8 = \sigma_8 (\Omega_\mathrm{m}/0.3)^{0.6}$. The red contours show the 68\%, 95\% and 99.9\% confidence contours of the fiducial analysis of KiDS-450 \citepalias{Hildebrandt2016}. The green contours show the 68\%, 95\% and 99.9\% confidence contours of baseline results ($\Lambda$CDM,TT,TE,EE+lowE+lensing) of Planck 2018 \cite{Planck2018_res}.
\label{fig:simgrid}}
\end{figure}

To generate the convergence maps used to train the CNN, we ran N-body simulations using the \Pkdgrav\ code \cite{Stadel2001}. \Pkdgrav\ uses the fast multipole expansion to accurately compute the forces between the particles in linear time and can be run with GPU support, accelerating the computation even further. \Pkdgrav\ is considered to be one of the most accurate and efficient N-body simulation codes and has been run successfully with more than a trillion particles \cite{Potter2017}.
Assuming a flat $\Lambda$CDM universe, we simulated a total of 57 different cosmologies in the $\Omega_\mathrm{m}-\sigma_8$ plane. The whole simulation grid is shown in figure \ref{fig:simgrid}. We fixed the remaining cosmological parameters to $\Omega_b = 0.0493$, $H_0 = 67.36$ and $n_s = 0.9649$ which corresponds to the baseline results ($\Lambda$CDM,TT,TE,EE+lowE+lensing) of Planck 2018 \cite{Planck2018_res}.
We chose the different cosmologies along the degeneracy parameter $\Sigma_8 = \sigma_8 (\Omega/0.3)^{0.6}$ inside the prior ranges $\Omega_\mathrm{m} \in [0.1, 0.5]$ and $\sigma_8 \in [0.4, 1.4]$, such that they cover the 3$\sigma$ confidence contours of the fiducial results of KiDS-450 \citepalias{Hildebrandt2016} and Planck 2018 \citep{Planck2018_res}. For each cosmology we ran a total of 12 simulations with 256$^3$ particles in a volume of $500^3$ Mpc$^3$. The initial conditions were generated at redshift $z_\mathrm{init} = 50$ using the initial condition generator \textsc{Music} \cite{MUSIC2011}. All simulations were run with 500 time steps to redshift $z_\mathrm{final} = 0.0$ and we used the default accuracy parameters $\theta = 0.4$, $\theta_{20} = 0.55$ and $\theta_2 = 0.7$. These parameters set the tree opening radius for different redshifts. \Pkdgrav\ divides the force calculation into particle-particle and particle-cell interactions. A high value of a $\theta$ parameter results in a smaller tree opening radius and less particle-particle interactions are calculated, while a small value of $\theta$ results in more particle-particle interactions. For high redshifts with a homogeneous particle distribution, smaller values of $\theta$ are necessary to reach the desired accuracy, as the force is not dominated by the surrounding, close particles. The parameter $\theta = 0.4$ was used for the redshift range between $z = 50$ and $z = 20$, $\theta_{20} = 0.55$ for the range between $z = 20$ and $z = 2$ and $\theta_2 = 0.7$ between $z = 2$ and $z = 0$. Each simulation generated a total of 51 snapshots covering the redshift range of the KiDS-450 galaxies from $z = 3.45$ to $z = 0.0$. For the early snapshots from $z= 3.45$ to $z = 1.55$ with less non-linear clustering we chose an output interval of $\Delta z = 0.1$. Afterwards, we reduced the output interval to $\Delta z = 0.05$ down to redshift $z = 0.0$. One should note that the time steps of \Pkdgrav\ are solely determined by the number of steps and the cosmological parameters used for the simulation. It was therefore not possible to output snapshots at the exact redshifts described above. Due to this, we decided to output a snapshot each time \Pkdgrav\ crossed a redshift of interest for the first time. The redshift difference between two output redshifts was therefore not constant and also slightly varying across the different cosmologies.

\subsubsection{Convergence Maps \label{subsec:conv_maps}}

As in previous work \cite{Fluri2018}, we used the \textsc{Ufalcon} package \cite{Sgier2019} to generate the convergence maps. \textsc{Ufalcon}  follows the procedure described in the appendix of \cite{Teyssier2009} and uses the Hierarchical Equal Area iso-Latitude Pixelization tool \citep{HEALPix} (\textsc{HEALPix}\footnote{Http://healpix.sourceforge.net.}). However, instead of generating full sky convergence maps, in this work we used the \emph{pencil beam} approach. Out of each snapshot at redshift $z_i$ we cut out a shell of particles with a thickness of $\Delta z = z_{i+1} - z_i$, where $z_{i+1}$ is the redshift of the previous snapshot. If necessary, we made use of the periodic boundary conditions of the simulations to achieve a patch size of 5x5 deg$^2$.
Afterwards, the particles were projected onto a regular grid using the gnomonic projection.
To increase the number of shells generated from one simulation, we repeated this process 1000 times for each snapshot. Each time we randomized the particle positions with random shifts, 90$^\circ$ rotations and parity flips. These shells were then used to generate convergence maps. A detailed description of the formalism can be found in \cite{Sgier2019, Fluri2018}.
The convergence of a given pixel $\theta_\mathrm{pix}$ can be calculated using the Born approximation in the following way
\begin{equation}
\small
\kappa(\theta_\mathrm{pix}) \approx \frac{3}{2}\Omega_\mathrm{m}\sum_bW_b\frac{H_0}{c}\int_{\Delta z_b}\frac{c\mathrm{d}z}{H_0E(z)}\delta\left(\frac{c}{H_0}\mathcal{D}(z)\hat{n}_\mathrm{pix},z\right), \label{eq:conv_map}
\end{equation}
where $\mathcal{D}(z)$ is the dimensionless comoving distance, $\hat{n}_\mathrm{pix}$ is a unit vector pointing to the pixels center and $E(z)$ is given by
\begin{equation}
\mathrm{d}\mathcal{D} = \frac{\mathrm{d}z}{E(z)}.
\end{equation}
The sum runs over all redshift shells and $\Delta z_b$ is the thickness of shell $b$.
Each shell gets the additional weight $W_b$ which depends on the redshift distribution of the source galaxies. For a given source redshift distribution, the weight is calculated using
\begin{equation}
W^{n(z)}_b = \frac{\int_{\Delta z_b}\frac{\mathrm{d}z}{E(z)}\int_z^{z_s}\mathrm{d}z'n(z')\frac{\mathcal{D}(z)\mathcal{D}(z,z')}{\mathcal{D}(z')}\frac{1}{a(z)}}{\int_{\Delta z_b}\frac{\mathrm{d}z}{E(z)}\int_{z_0}^{z_s}\mathrm{d}z'n(z')},
\end{equation}
with redshift boundaries $z_0 = 0.0$ and $z_s = 3.45$, covering the whole range of the KiDS-450 catalog. For each of the four redshift bins we used the effective source redshift distribution obtained with the weighted direct calibration (DIR), which is the default calibration method in \citetalias{Hildebrandt2016}. This method was suggested by \cite{Lima2008} and first applied by \cite{Bonnett2016}. It uses a k-nearest neighbor search to estimate the volume density of objects in the multi-dimensional magnitude space in photometric and spectroscopic catalogs. These estimates, together with a re-weighting scheme, are then used to infer the effective redshift distribution of the photometric catalog. A detailed description of the parameter choices can be found in \citetalias{Hildebrandt2016}.

To check the accuracy of our simulations we compared the power spectrum of the convergence maps with theoretical predictions.
The theoretical predictions were obtained using the \textsc{Nicaea} code \cite{nicaea2009}.
This comparison is shown in figure \ref{fig:spectra}.
We find that our simulations are able to accurately describe the matter distribution of the universe up to a multipole of $\ell \sim 3000$.
However, in order to reduce the impact of smaller scales in our analysis, smoothed maps that were fed into the CNN with a Gaussian kernel of width $\sigma_s = 2.34$ arcmin, which is equivalent to the pixel scale.
The same smoothing was applied to the maps used for the power spectrum analysis.

With this approach, we generated 12'000 convergence maps for each of the simulated cosmologies. We used 10'000 maps generated with 10 independent simulations to train the CNN and reserved 2'000 as the test set.
The test set was solely made out of maps created using simulation boxes that were not used to make the training set, as a precaution against the CNN learning features of the randomization procedure explained above.

\subsubsection{Intrinsic Alignment \label{subsec:IA}}

\begin{figure*}
\includegraphics[width=1.0\textwidth]{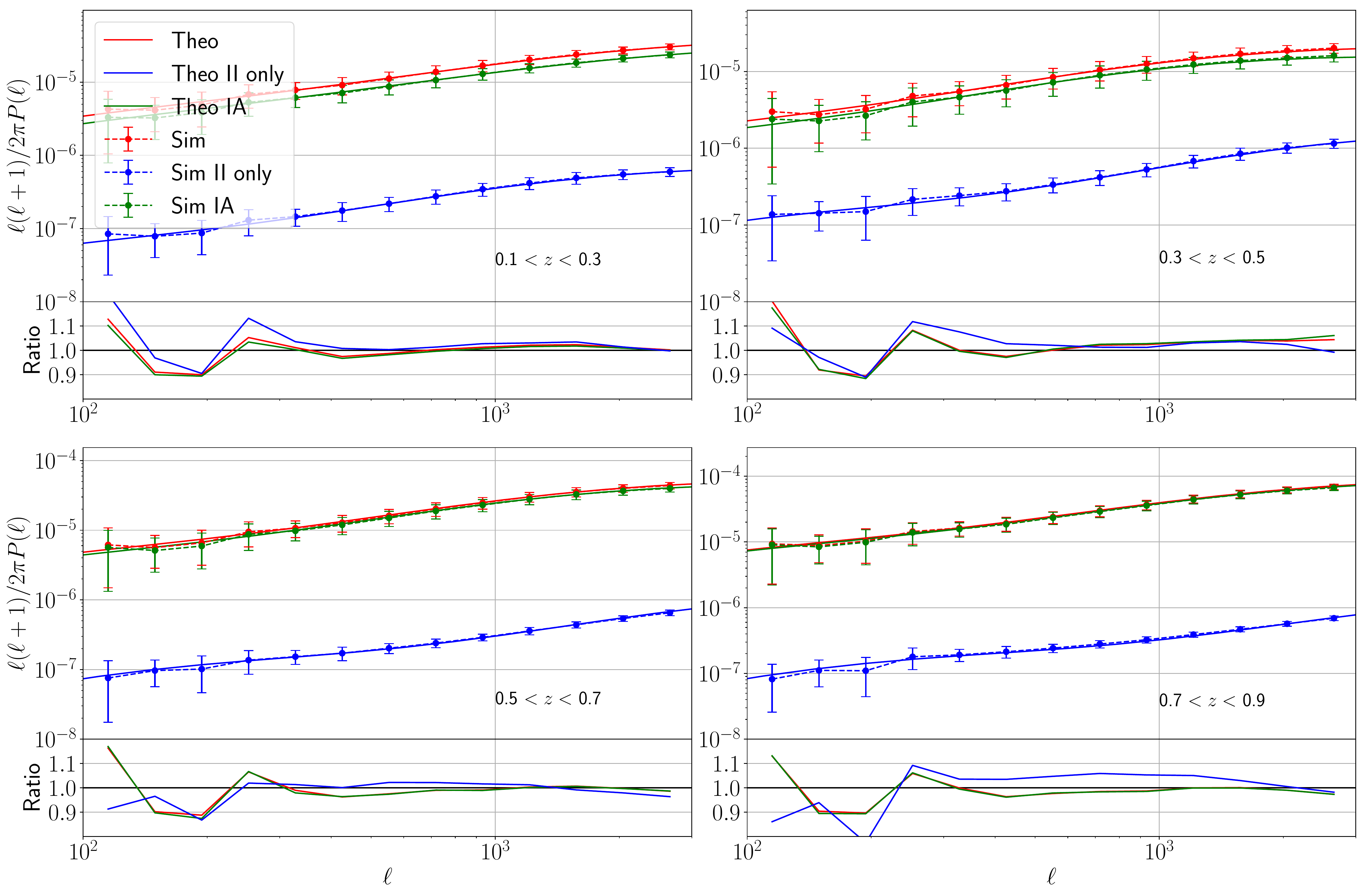}
\caption{The average power spectrum of 100 convergence maps (10 per simulation in our training set, no masking, no noise) for each redshift bin and with intrinsic alignment ($A_\mathrm{IA} = 1$) for a cosmology with $\Omega_\mathrm{m} = 0.29$ and $\sigma_8 = 0.78$. The theoretical predictions were obtained using the \textsc{Nicaea} code \cite{nicaea2009}. The maps were generated with an increased resolution of 256$\times$256 pixels to avoid power loss through pixelization effects. \label{fig:spectra}}
\end{figure*}

Intrinsic galaxy alignment is one of the most important systematic effects in weak gravitational lensing. There are two types of intrinsic alignment: the intrinsic-intrinsic (`II') type is caused by massive objects that align the intrinsic ellipticities of galaxies around them, and gravitational-intrinsic (`GI') type concerns the correlations between the intrinsic ellipticity of a foreground galaxy and the shear of a background galaxy.
To model the intrinsic alignment in our simulated convergence maps we used the model developed by \cite{Hirata2004, Bridle2007, Joachimi2011}.
This model has also been used in the fiducial results of the KiDS-450 analysis \citepalias{Hildebrandt2016} and many other weak lensing surveys. Originally, it was developed to predict the effects of intrinsic alignment on the power spectrum. It connects the intrinsic alignment power spectrum with the non-linear matter power spectrum
\begin{align}
P_\mathrm{II} &= F(z)^2P_\delta(k, z) \label{eq:IAII} \\
P_\mathrm{GI} &= F(z)P_\delta(k, z), \label{eq:IAGI}
\end{align}
where $F(z)$ is a cosmology and redshift dependent term
\begin{equation}
F(z) = -A_\mathrm{IA}C_1\rho_\mathrm{crit}\frac{\Omega_\mathrm{m}}{D_+(z)}\left(\frac{1 + z}{1 + z_0}\right)^\eta\left(\frac{\bar{L}}{L_0}\right)^\beta,
\end{equation}
where the intrinsic alignment amplitude $A_\mathrm{IA}$, $\eta$ and $\beta$ are the free parameters of this model, ${C_1 = 5 \times 10^{-14} h^{-2}M_\odot\mathrm{Mpc}^3}$ is a normalization constant, $\rho_\mathrm{crit}$ is the critical density at $z=0$ and $D_+(z)$ normalized linear growth factor, so that $D_+(0) = 1$. The parameters $L_0$ and $z_0$ are arbitrary pivot points to model the redshift and luminosity dependence. As in \citetalias{Hildebrandt2016} we fixed $\eta = \beta = 0$ and did not consider the redshift and luminosity dependent terms. 

We used equations \eqref{eq:IAII} and \eqref{eq:IAGI} to implement the model on convergence map level, such that it reproduces the results for the power spectrum. Similar to equation \eqref{eq:conv_map} one can calculate the intrinsic alignment part of the convergence map
\begin{equation}
\kappa_\mathrm{IA} \approx \sum_bW_b^\mathrm{IA}\frac{H_0}{c}\int_{\Delta z_b}\frac{c\mathrm{d}z}{H_0E(z)}\delta\left(\frac{c}{H_0}\mathcal{D}(z)\hat{n}_\mathrm{pix},z\right),
\end{equation}
where the weights are given by
\begin{equation}
W_b^\mathrm{IA} = \frac{\int_{\Delta z_b}\mathrm{d}zF(z)n(z)}{\left(\int_{\Delta z_b}\frac{\mathrm{d}z}{E(z)}\int_{z_0}^{z_s}\mathrm{d}z'n(z')\right)}
\end{equation}
The accuracy of this model is shown in figure \ref{fig:spectra}.
For the intrinsic alignment amplitude we chose the same non-informative prior $-6 < A_\mathrm{IA} < 6$ as in \citetalias{Hildebrandt2016}. It is important to note that $\kappa_\mathrm{IA}$ depends linearly on the intrinsic alignment amplitude $A_\mathrm{IA}$. Due to this, it is only necessary to compute a single intrinsic alignment map for each corresponding convergence map. A convergence map with any intrinsic alignment amplitude can then be generated by simply adding the two maps with the appropriate factor.

\subsubsection{Shape and Measurement Noise \label{subsec:noise}}

In order to train the CNN with the right noise properties we used the positions and ellipticity magnitudes from the galaxy catalog.
The noise was generated by randomly rotating each galaxy
\begin{equation}
\gamma_\mathrm{pix}^\mathrm{noise} = \frac{\sum_{j \in \mathrm{pix}}\exp(\theta_j i)w_j (\gamma_j - c_z)}{\sum_{j \in \mathrm{pix}}w_j},
\end{equation}
where the $\theta_j$ were drawn uniformly from $[0, 2\pi)$.
We implemented this procedure on GPUs to compensate the computational costs of this approach.
This also allowed a ``on the fly'' generation of noise maps during the training of the CNN.
This approach preserves the spatial variation of the number density of galaxies and the shape noise amplitude across the footprint.

\subsubsection{KS Inversion \label{subsec:KSinversion}}

The KS inversion \cite{KS1993} allows for transforming convergence maps into shear maps and vice-versa.
We used it to generate shear maps from simulated convergence, as well as to generate convergence from shear maps, after having added the systematic effects.
By making use of the small angle approximation and the fast Fourier transform (FFT) one can relate the shear $\gamma$ and the convergence $\kappa$ via the lensing potential $\Psi$, leading to the following relation in Fourier space
\begin{equation}
\tilde{\gamma}_i = D_i\tilde{\kappa}, \label{eq:kappa_shear}
\end{equation}
where $\tilde{\gamma}_i$ represents the Fourier-transformed real or imaginary part of the complex shear $\gamma$, $\tilde{\kappa}$ is the Fourier transform of the convergence $\kappa$ and the kernel is given by
\begin{equation}
\begin{pmatrix}
D_1 \\
D_2
\end{pmatrix}
=
\begin{pmatrix}
\frac{l_1^2 - l_2^2}{l^2} \\
\frac{2l_1l_2}{l^2}
\end{pmatrix}
\end{equation}
where $(l_1, l_2)$ is the 2D Fourier conjugate of the position $(\theta_1, \theta_2)$ in real space and $l^2 = l_1^2 + l_2^2$. It is important to note that the kernels $D_i$ each vanish for certain values of $(l_1, l_2)$. This issue can be alleviated for the transformation from shear maps to convergence maps by making use of the relation
\begin{equation}
\sum_i \vert D_i \vert^2 = 1, \label{eq:kernel_rel}
\end{equation}
which was obtained by \cite{Kaiser1992}. Combining equations \eqref{eq:kappa_shear} and \eqref{eq:kernel_rel} one can express the Fourier transform of the convergence in the following way
\begin{equation}
\tilde{\kappa} = \sum_i D_i \tilde{\gamma}_i.
\end{equation}
Afterwards one can simply perform an inverse FFT to obtain the actual convergence.
However, for the transformation from convergence maps to shear maps, it is not possible to recover the modes missing due to the vanishing kernels. Since the examined CNN works on shear maps, it is important to address this issue.
For the simulations used for the training of the CNN, some information is lost during the transformation from convergence to shear.
However, this procedure would not be used for the survey data.
We, therefore, removed these missing modes from the noise maps that were added to the simulated shear maps, as well as from the survey data before the final inference.
This was done by performing a FFT on these maps and setting all modes, where the corresponding kernels vanish, to zero, and then transforming them back into real space.
The FFT to remove the modes was performed on masked maps, while the KS inversion of the simulated convergence maps was done before the observational mask was applied. Therefore, this ``mode removal" procedure could potentially introduce a bias in the inference pipeline. 
We examined this effect by using ray-traced simulations (see section \ref{subsec:raytrace}). Ray-tracing allows for generating equivalent convergence and shear maps directly from the produced particle shells. The ``mode removal" procedure can then be compared to the standard KS inversion of the convergence maps.
The results of this comparison are shown in section \ref{sec:effects_sys}.

Another common problem with the KS inversion is that the FFT assumes periodic boundary conditions. Using the KS inversion on non-periodic data can lead to boundary effects. Further, masking of the data can also lead to similar effects.
To minimize these effects we applied zero-padding to with 64 pixels on each side before performing the FFT.
This was done for each simulated convergence map that was transformed, as well as whenever the ``mode removal'' procedure was applied.

\subsection{Systematic Effects \label{sec:systematics}}

In this section, we give an overview of all systematic effects considered in our analysis.
The considered systematic effects include: shear multiplicative bias, photometric redshift error, baryon effects, super survey modes, simulations engine and resolution, and shear projection methods.
The effects of the shear multiplicative bias and photometric errors are marginalized in the likelihood analysis.
We test the impact on $S_8$ of the remaining effects by comparing the difference between the fiducial mock data with mock data that includes the systematic effect of interest.
To isolate the desired effect, we kept the simulations seeds the same, wherever possible.
The simulations that include the systematic effects were not used for training the network.
The results of these tests are presented in section \ref{sec:effects_sys}.
Our fiducial mock observation was generated from a simulation with the values $\Omega_\mathrm{m} = 0.25$ and $\sigma_8 = 0.849$, as measured from KiDS-450 by \citetalias{Hildebrandt2016}.

\subsubsection{Photometric Redshift}

We used the 1'000 publicly available bootstrapped `DIR' redshift distributions to incorporate the uncertainty of the effective source redshift distribution in our analysis. However, due to memory limitations, it was not possible to generate each convergence map in the training and test sets with all of the 1'000 effective source redshift distributions. Therefore, the convergence maps used for training were generated with the mean `DIR' distribution. The bootstrapped versions were then only used to obtain the summary statistics of the test set used for the inference (see section \ref{sec:inference}), such that our likelihood analysis marginalized over this error.

\begin{figure*}
\includegraphics[width=1.0\textwidth]{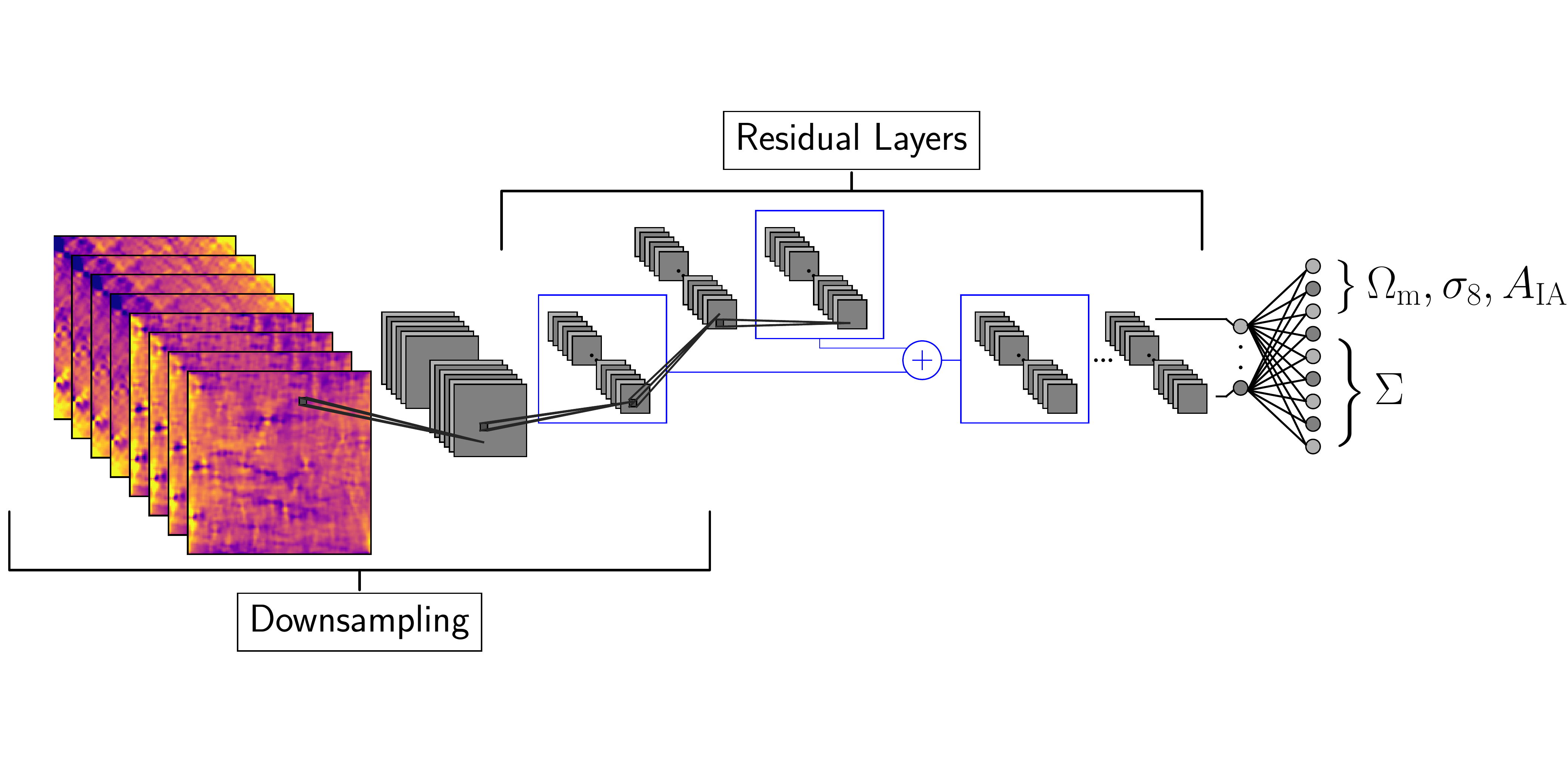}
\caption{Sketch of the network architecture: The input is downsampled twice which is followed by the residual blocks (see figure \ref{fig:resblock}) and the final output layer. We examined architectures with 10, 15 and 25 residual blocks. Three values of the output are used as predictions for $\Omega_\mathrm{m}$, $\sigma_8$ and $A_\mathrm{IA}$, the other six are used to build the covariance matrix of the loss in equation \eqref{eq:likelihoodloss}.
\label{fig:network_sketch}}
\end{figure*}

\subsubsection{Multiplicative and Additive Shear Bias \label{subsec:shearbias}}

As in \citetalias{Hildebrandt2016} we followed \cite{Refregier2006a} and parametrized the shear bias as
\begin{equation}
\gamma^\mathrm{obs} = (1 + m_b)\gamma^\mathrm{true} + c,
\end{equation}
where $m_b$ is the multiplicative shear bias and $c$ the additive shear bias term. A detailed description of the calibration method of the KiDS-450 fiducial analysis \citetalias{Hildebrandt2016} can be found in \cite{Conti2017}. Similar to \citetalias{Hildebrandt2016} we calibrated the additive shear bias directly from the observed data for each redshift bin and each of our 20 projected patches and neglect its uncertainty. This additive bias was corrected per shear component and averaged over all patches and both shear components we found $c = [$-0.20,  4.53,  2.26, -0.36$]\times10^{-4}$ for our four redshift bins. These values were then subtracted from each galaxy (see eq. \eqref{eq:shear}). Further, one should note that we obtained slightly different values than the KiDS-450 fiducial analysis \citetalias{Hildebrandt2016}, since we were only using a subset of the whole catalog.

Further, we estimated the multiplicative shear bias per redshift bin using the publicly available, per-object estimates of the KiDS-450 observed data, by calculating their weighted average
\begin{equation}
m_b^z = \frac{\sum_{i}m_b^iw_i}{\sum_{i}w_i},
\end{equation}
where the sum runs over all used galaxies in a given redshift bin and $m_b^i$ is the estimated multiplicative correction of the galaxy. The publicly available estimates were obtained using state-of-the-art simulations and the detailed procedure can be found in \cite{Conti2017}. We found $m_b^z = [$-0.015,  -0.011,  -0.01, -0.02] for our four redshift bins and, as in \citetalias{Hildebrandt2016}, we assume a combined statistical and systematic uncertainty of $\sigma = 0.01$ for each value. These values are consistent with the method used in \citetalias{Hildebrandt2016}, as it was shown in \cite{Conti2017}. Further, since KiDS-450 power spectrum analysis \cite{KIDSpower} also followed \cite{Conti2017} to estimate the multiplicative bias, we expect them to be consistent as well.

The additive shear bias was removed from the observed data before generating the noise maps. The multiplicative correction was applied on the simulated convergence maps rather than on the observed data, by multiplying the simulated maps with the appropriate factor. A detailed description of how we applied this correction during the training and the inference is provided in sections \ref{subsec:network} and \ref{sec:inference}.

\subsubsection{Baryonic Effects}

Baryonic effects can significantly change the weak lensing signal on small scales \cite{Semboloni2011, Mohammed2014, Osato2015, Schneider2018}.
The main KiDS-450 analysis used the \textsc{HMcode} model \cite{Mead2015} with a prior of $B \in [2, 4]$, where $B$ is the parameter regulating the strength of baryonic effects, with $B=3.13$ corresponding to no baryonic effects.
For that analysis, marginalizing over this prior led to $\approx$20\% increase in the size of the $S_8$ constraints.

Adding the effects of baryons to the dark matter only simulations is a challenging task, and only recently approaches have emerged to tackle this problem:
\cite{Schneider2018} developed a method to modify the positions of particles using a parametric model, while \citep{baryonpainting} used deep learning to paint the Baryon effects on the lensing maps based on hydrodynamic simulations.
However, adding baryonic effects would result in several additional nuisance parameters making the problem substantially more complex.
That is why we decided to neglect baryonic effects in our fiducial analysis configuration.
We did, however, test the impact of baryons with the help of mock observations that include these effects on the map level.
This was achieved by applying the \textit{baryon correction model} \cite{Schneider:2015wta, Schneider2018}, which is a method to modify the density field of gravity-only N-body simulations in order to mimic baryonic feedback effects.
In more detail, the baryonic correction model displaces particles around N-body halos, in order to obtain more realistic halo profiles that include effects from star formation and AGN feedback.
The halos were identified using the \textsc{Amiga} halo finder \cite{Gill2004, Knollmann2009}.
The baryonic correction model has been shown to be in good agreement with full hydrodynamical simulations \cite{Schneider2018}.

In order to account for the inherent uncertainties related to the baryonic feedback processes, we created three mock observations with weak, best-guess, and strong baryonic corrections. These three models correspond to the average benchmark cases A, B and C described in table 2 of \cite{Schneider2018}. They are motivated by observed X-ray gas fractions \cite{Sun2009, Vikhlinin2009, Gonzalez2013} including current uncertainties of the hydrostatic mass bias \cite{Eckert2016}. The impact of these three benchmark models on the resulting weak-lensing constraints is described in section \ref{sec:effects_sys}.

\subsubsection{Super Survey Modes and Finite Box Effects}

The periodic boundary conditions used in N-body simulations can introduce a bias caused by finite box effects. The large-modes and their coupling to small-modes are usually not properly resolved which can lead to underestimated errors. To test that our pipeline is robust to this effect we generated a mock observation from a simulation where we doubled the box size to $L = 1000$Mpc and the particle number $N_\mathrm{part} = 512^3$.

\subsubsection{Simulation Settings}

Our inference pipeline needs to be robust with respect to the simulation software used to generate the training data and an increased number of particles inside the simulations. To verify the robustness regarding the simulation software, we generated a mock observation with the TreePM-code \textsc{Gadget-2} \cite{Springel2005}. One advantage of Gadget-2 is that it can output snapshots at specified redshifts. We, therefore, chose the redshifts described in section \ref{sec:sims} and used the same initial conditions as for our fiducial mock observation.

The pipeline also needs to be robust to the number of particles used in the simulations.
To test this, we generated a simulation equivalent to the one used for our fiducial mock observation, but with the number of particles increased to $512^3$.

\subsubsection{The Born Approximation \label{subsec:raytrace}}

The Born approximation we used to generate the convergence maps is a first-order approximation of the convergence. This approximation can have significant effects on the resulting constraints \cite{Petri2017}. To test the sensitivity of our pipeline to this approximation, we created a mock observation that was fully ray-traced. This was done using the \textsc{Lenstools} package \cite{Petri2016}. \textsc{Lenstools} provides a multi-plane ray-trace algorithm that solves the lensing equation of the different mass shells in Fourier space. These mass shells should have a much higher resolution than the final convergence maps to avoid pixelization effects. We, therefore, used our simulation with an increased number of particles described above to generate mass shells with a resolution of 4096 pixels per side. Afterwards, we computed the shear of 2048 galaxies per side distributed according to the DIR redshift distributions to generate convergence and shear mock observations.

\subsection{Convolutional Neural Network \label{subsec:network}}

\subsubsection{Architecture}

\begin{figure}
\centering
\includegraphics[width=0.2\textwidth]{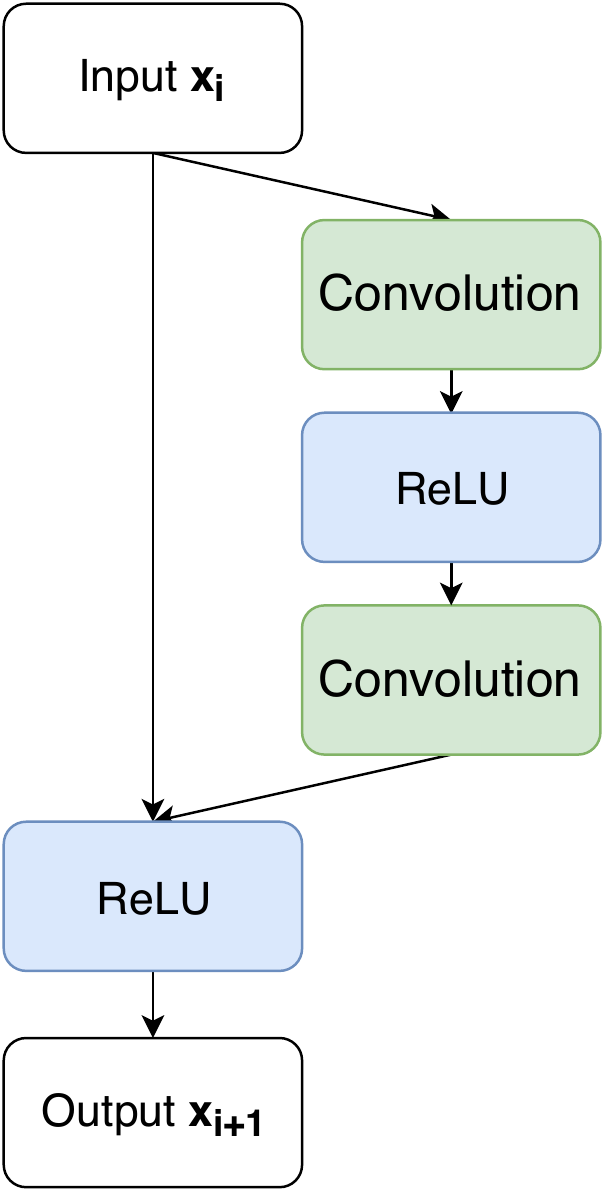}
\caption{A residual layer used in the networks. The input is convolved with a weight kernel that does not change the channel dimension and padding is applied such that the image dimension is conserved. This is followed by an activation function. This process is then repeated and the original input is added to the output before the final activation. We do not apply batch normalization in our residual blocks as it did not improve our results. \label{fig:resblock}}
\end{figure}

We examined three different architectures of convolutional neural networks. The implementation was done using \textsc{Tensorflow} \cite{tensorflow2015}. All of them were built out of residual layers \cite{He2015} shown in figure \ref{fig:resblock} and only the filter size and depth were different. A sketch of this type of architectures is shown in figure \ref{fig:network_sketch}.
The first two layers of the networks are convolutional layers with a stride of two, a linear rectified unit (ReLU) activation function and no padding to down-sample the input. The first layer was set to increase the number of channels to 128, all other convolutional layers were set to conserve this channel dimension. We examined one network with a filter size of 5$\times$5 and 10 residual layers as described in figure \ref{fig:resblock}. Further, we examined two networks with a filter size of 3$\times$3 using 15 and 25 residual blocks. The last layer is always a fully connected layer with no activation function to map the output of the last residual block to the desired output size. Similarly to \cite{Fluri2018, Levasseur2017ApJ}, we used a negative log-likelihood loss function
\begin{equation}
L = \frac{1}{2}\left(\ln\left(\left\vert\Sigma\right\vert\right) + \left(\theta_p - \theta_t\right)^T\Sigma^{-1}\left(\theta_p - \theta_t\right) \right), \label{eq:likelihoodloss}
\end{equation}
where $\theta_p$ represents the vector of predicted parameters, $\Sigma^{-1}$ the predicted inverse covariance matrix and $\theta_t$ are the true parameters. The network was trained on the two cosmological parameters $\Omega_\mathrm{m}$ and $\sigma_8$, and the intrinsic alignment amplitude $A_\mathrm{IA}$. To have all parameters in approximately the same dynamic range we divided the intrinsic alignment amplitude by a factor of ten, such that the network was trained on $\theta_t = (\Omega_\mathrm{m}^t, \sigma_8^t, A_\mathrm{IA}^t/10)$. Further, it is important that the network always predicts a valid covariance matrix. This was done by using the fact that every positive-definite symmetric matrix has a Cholesky decomposition of the form
\begin{equation}
\Sigma^{-1} = LL^T,
\end{equation}
where $L$ is a lower triangular matrix. We, therefore, let the network predict the six free parameters of this lower triangular matrix instead of the covariance matrix directly. The inverse covariance matrix and its determinant were then built using this decomposition.
Unlike in \cite{Fluri2018}, we did not use the regularization term.
The bias of the network prediction is handled in the likelihood function, as described in \ref{sec:inference}.

\subsubsection{Input Pipeline}

A schematic diagram of the input pipeline is shown in figure \ref{fig:pipeline}. For a single cosmology, the pipeline started by randomly selecting 20$\times$4 (20 patches, 4 redshift bins) convergence maps and corresponding intrinsic alignment maps out of our training set.  These maps were then added with an intrinsic alignment amplitude $A_\mathrm{IA}$ drawn from a uniform distribution according to our prior $A_\mathrm{IA} \in [-6, 6]$. Afterwards we implemented the multiplicative shear bias correction by multiplying all the maps from a given redshift bin with its corresponding bias term $(1 + m_b^z + \epsilon_\mathrm{err})$, where $\epsilon_\mathrm{err}$ was drawn from a normal distribution with mean $\mu = 0$ and standard deviation $\sigma = 0.01$ to model the uncertainty of our multiplicative shear bias estimate. The resulting maps were padded with 64 pixels of value zero on each side and transformed into shear maps using the KS inversion. We then masked each of the 20$\times$8 (two shear maps per redshift bin) patches with the observational mask obtained from the survey data. Parallel to that process, we generated a random noise map from the observed data as explained in section \ref{subsec:noise}. After applying our ``mode removal" procedure (see \ref{subsec:KSinversion}) we added the generated noise maps to our simulated shear maps and smooth each map individually with a Gaussian smoothing kernel with $\sigma_s = 2.34$ arcmin, equivalent to the pixel size. At this point the input for a single cosmology had the dimension (20, 128, 128, 8), corresponding to the 20 projected patches, the 4 different redshift bins times the 2 shear components and the 128$\times$128 pixels of each map. However, we found the best results by additionally providing the network an estimate of the convergence. We, therefore, applied a second KS inversion and concatenated all maps to an input with dimensionality (20, 128, 128, 12). The last KS inversion was performed solely to improve the convergence of the network and we did not use any padding, because all considered systematic effects were already added to the maps. The network predicts a set of parameters $\theta_p = (\Omega_\mathrm{m}^p, \sigma_8^p, A_\mathrm{IA}^p)$, as well as the corresponding errors, for each of the 20 patches individually.

\subsubsection{Training}

The training was performed with the \texttt{Adam} optimizer \cite{Adam2014} with first and second-moment exponential decay rates equal to 0.9 and 0.999 respectively and an the initial learning rate equal to $0.5\times 10^{-5}$. Further, we applied gradient clipping and normalized the length of the gradient of the weights to 100 if the global norm would exceed this value. To reduce the training time, we trained the networks asynchronously on 16 GPUs for 1'500'000 iterations where we fed the network maps from eight cosmological parameter combinations at each iteration. For the network architectures with 10 and 15 residual blocks, it was not necessary to slowly increase the smoothing scale and noise level as in our previous work \cite{Fluri2018}. However, the network with 25 residual blocks did not converge when trained directly on noisy, smoothed maps. To address this, we pre-trained this network for 1'000'000 iterations on noise-free, smoothed maps, before training it on noisy maps.

In the asynchronous training approach, the weights of the network are governed by parameter servers and distributed to each GPU. Each GPU computes the gradients of a single mini-batch (input of eight cosmologies) and sends them back to the parameter server to update the weights. The 16 worker GPUs update the weights of the network asynchronously and as fast as possible. While being extremely efficient computationally, this approach can lead to slightly worse results than synchronous training on a single GPU. To alleviate this problem we trained each network for 100'000 iterations on a single GPU after the completion of the asynchronous training. However, we did not find a significant difference in the performance of the networks after the synchronous training.

We evaluated one batch of our test set every 250 iterations. An example of such a loss is shown in appendix \ref{ap:loss}. We did not observe any signs of overfitting for the examined architectures.

\section{Inference \label{sec:inference}}

To generate cosmological constraints using the predictions of the networks we followed the same approach as in previous work \cite{Fluri2018}. We used the predictions of our test set as summary statistics and perform a standard likelihood analysis. An example of the predictions of a single architecture is presented in appendix \ref{ap:evals}. To perform this evaluations we randomly split the generated mass shells of our test set into 100 sets of 20 patches. For each set, we generated convergence maps with a randomly drawn redshift distribution from the 1'000 bootstrapped, publicly available `DIR' redshift distributions to marginalize over the photometric redshift error. We then fed these maps into our inference pipeline with randomly drawn multiplicative shear bias error estimates and noise maps. The predictions of each set of 20 patches were then averaged to obtain the final prediction of a simulated survey. Afterwards, we repeated the process with new random seeds to obtain a total of 200 predictions of simulated surveys for each of our 57 cosmologies and for 25 linearly spaced intrinsic alignment amplitudes $A_\mathrm{IA} \in [-6,6]$. Finally, we used these 200 predictions to calculate the mean $\hat{\mu}$ and covariance matrix $\hat{S}$. These mean predictions and covariance matrices were then linearly interpolated across the grid to perform a likelihood analysis. We used the Gaussian likelihood described in \cite{Jeffrey2018} that is based on \cite{Sellentin2015} and includes the uncertainty of the estimated means and covariance matrices. The likelihood of a given measurement $d$, given a covariance matrix $\hat{S}$ estimate that is obtained from $N$ simulations and an estimated mean $\hat{\mu}$ from $M$ simulations is given by
\begin{equation}
P(d \vert \hat{\mu}, M, \hat{S}, N) \propto \vert\hat{S}\vert^{-\frac{1}{2}}\left( 1 + \frac{M}{(M+1)(N-1)}\mathbf{Q}  \right)^{-\frac{N}{2}},
\end{equation}
where
\begin{equation}
\mathbf{Q} = \left(d - \hat{\mu}\right)^T\hat{S}\left(d - \hat{\mu}\right),
\end{equation}
and in our case we had $M = N = 200$.

\subsection{Combining networks}
\label{sec:combining_networks}

While the results of a single network were already promising, we found that combining the prediction of the different architectures improved the results even further. We combined the different ResNet architectures by concatenating their predictions into a long data vector $d = (d_1, d_2, d_3, d'_1, d'_2, d'_3)$, where the $d_i$ are the predictions of $\Omega_\mathrm{m}$, $\sigma_8$ and $A_\mathrm{IA}$ of a single network after the asynchronous training is finished and the $d'_i$ are the predictions of the different architectures after the synchronous training. This lead to an output vector of length 18 used in our fiducial analysis. The cosmological constraints from our fiducial mock observation obtained with the combination from the different architectures and from a single architecture with 10 residual blocks are shown in figure \ref{fig:fiducial_cons} in appendix \ref{sec:multi_vs_single_cnn}.
The covariance matrix of this concatenated vector had a size of $18\times18$, taking into account all correlations between the output of the networks and is interpolated as described in section \ref{sec:inference}.
It is important to note that the linear interpolation used in our likelihood analysis is only valid inside the convex hull of our simulated cosmologies. This hull, therefore, defined our priors on the cosmological parameters. The combination of the different architectures reduces the area of the 95\% confidence contours by approximately 14\%.
This improvement shows that a single network was not able to fully extract the information from the weak lensing maps and combining networks with different architectures can bring us closer to that goal.
One reason for this is potentially the high noise levels which make it difficult for a single network to converge.

\begin{figure*}
\includegraphics[width=1.0\textwidth]{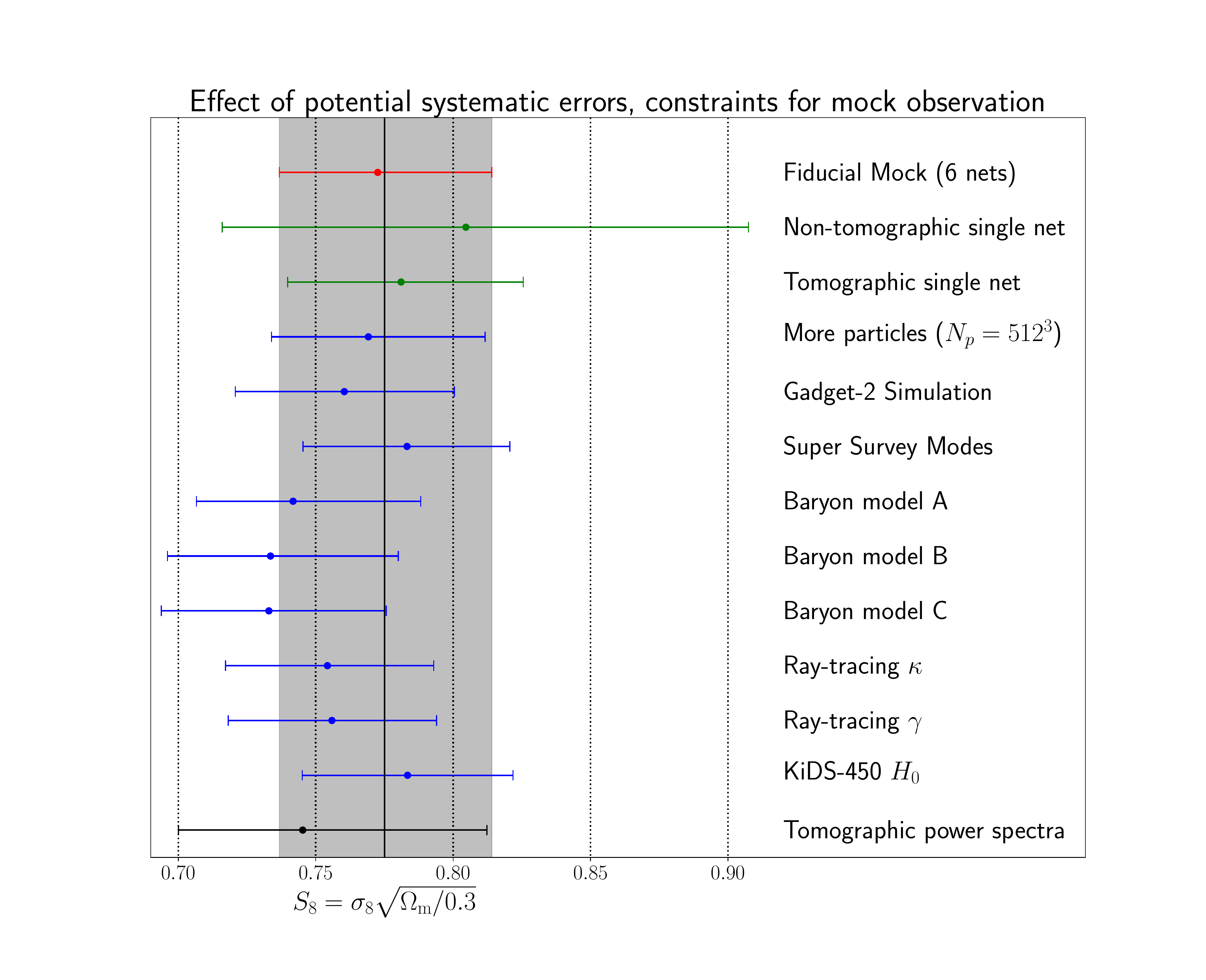}
\caption{
Impact of various potential systematic effects on the $S_8$ constraints.
The blue constraints correspond to different mock observations obtained with our fiducial inference pipeline.
The first three constraints correspond to the mock observation with identical random seeds, but with different inference pipeline settings.
The black constraints were obtained using a tomographic power spectrum analysis. The black solid line corresponds to the true $S_8$ value of the mock observation.
 \label{fig:systematics}}
\end{figure*}

\subsection{Power spectrum Analysis}

We performed a power spectrum (PS) analysis in a similar way to the CNN analysis in order to assess the advantage of the CNN in a tomographic setting.
The likelihood for the PS analysis was constructed using our simulation test set as theoretical predictions, similar to our previous work \cite{Fluri2018}.
This was done by computing the auto- and cross-spectrum of all our patches in our test set.
The prediction of a single simulated survey was then obtained by again averaging over our 20 patches.
The auto- and cross-spectrum are calculated in seven logarithmic multi-pole bins in the range $\ell \in [75, 3500]$.
The maps used for this analysis were identical to the ones used by the CNN, including the smoothing scale of $\sigma=2.34$ arcmin.
This smoothing scale reduces the power spectrum to 10\% of its original values at $\ell\sim2200$.
We did not interpolate the covariance matrices in our power spectrum analysis and use a fixed covariance matrix from our test set. The cosmological parameters of the covariance matrix were chosen such that they were close to the parameters of our mock observation and we set $A_\mathrm{IA} = 0$.
We verified that the CNN and PS give consistent result on simulations.
The generated constraints on the degeneracy parameter $S_8 = \sigma_8\sqrt{\Omega_\mathrm{m}/0.3}$ are shown in figure \ref{fig:systematics}.
More details of our power spectrum analysis can be found in appendix \ref{ap:specs}.

\section{Effects of the Systematic Errors \label{sec:effects_sys}}

The results of our pipeline tests are shown in figure \ref{fig:systematics}. All of the constraints were obtained by using the average prediction of all simulated surveys from a simulation in a given setting as a measurement vector $d$.
The number of simulated surveys was 40 for N-body simulations with 256$^3$ particles and, because of the increased computational costs, 8 for mock observations generated from N-body simulations with 512$^3$ particles. This was done to decrease the statistical error in the size and position of the constraints. All mock observations had an intrinsic alignment amplitude of $A_\mathrm{IA} = 0$.
Wherever possible we used the same seeds to generate the initial conditions for the simulations.

\begin{figure*}
\subfigure{
\includegraphics[width=0.5\textwidth]{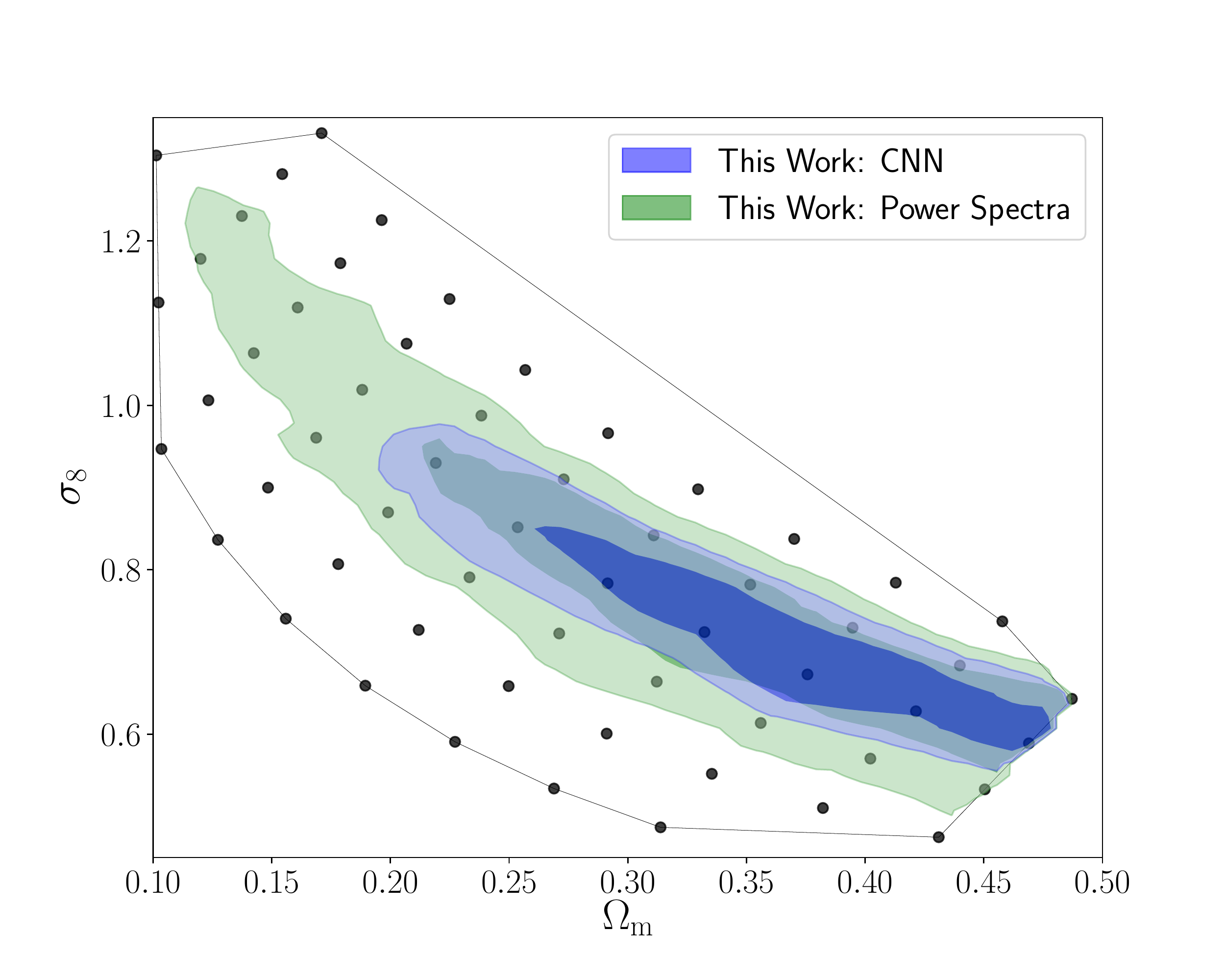}
}%
\subfigure{
\includegraphics[width=0.5\textwidth]{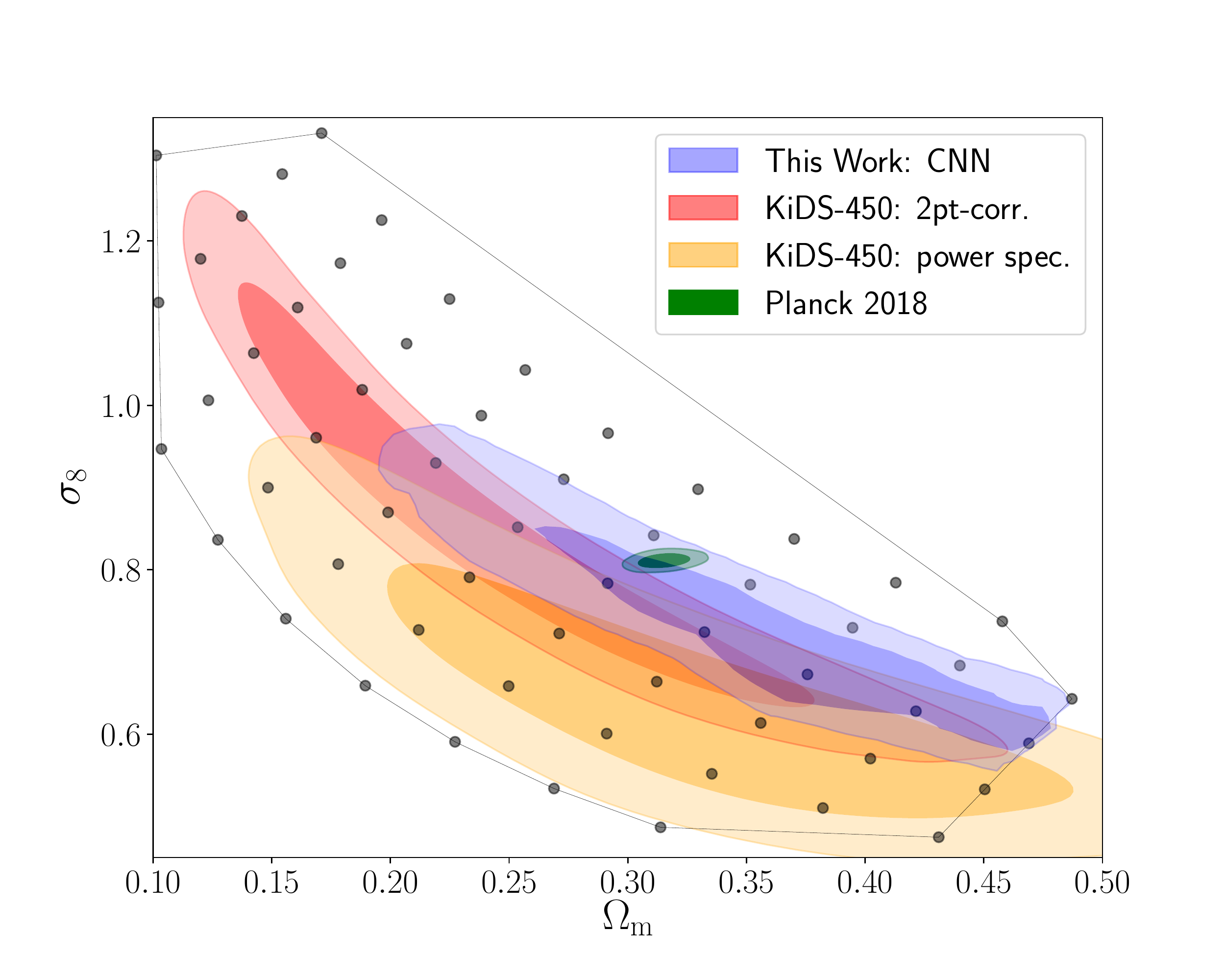}
}%
\caption{Left Panel: The 68\% and 95\% confidence contours from our tomographic power spectrum analysis and the constraints obtained with the CNN. Right Panel: The 68\% and 95\% confidence contours from our CNN, the fiducial results from KiDS-450 \citepalias{Hildebrandt2016}, the KiDS-450 power spectrum analysis with 3 redshift bins \cite{KIDSpower} and the baseline results from Planck 2018 \cite{Planck2018_res}. The convex hull of our simulated cosmologies (prior) is drawn in black. \label{fig:CNN_vs_other}}
\end{figure*}

The constraints from a single architecture are obtained with the parameters of the network with 10 residual blocks after the asynchronous training. For completeness, we also show the constraints generated from a non-tomographic analysis. Tomography improves the constraints on $S_8$ by almost 60\%. This large improvement is mostly caused by the fact that a tomographic analysis is able to break the degeneracy of the intrinsic alignment amplitude and the cosmological parameters (see appendix \ref{ap:IA_amp} for more details).

The different simulations setting do not have a significant impact on the constraints. Increasing the number of particles leaves the results almost unchanged, while the Gadget-2 simulation prefers a slightly lower $S_8$ value.

The two versions of the ray-traced mock observation are consistent, meaning that a possible bias from our ``model removal'' procedure (see section \ref{subsec:KSinversion}) is negligible. The two mock observations were obtained using the multi-plane ray-trace algorithm, which made it possible to produce equivalent convergence and shear maps directly from the Jacobians and not relying on any inversion methods. The shear maps (``Ray-tracing $\gamma$'') were fed into our inference pipeline exactly the same way as we handle the observed KiDS-450 data, while the convergence maps (``Ray-tracing $\kappa$'') were fed into our inference pipeline in the same way as the training maps.
However, it is evident that the ray-tracing itself leads to sightly smaller values of $S_8$, meaning that the used Born approximation introduces a small systematic bias. We leave this bias to future work, as ray-tracing is computationally much more intensive than the Born approximation and usually also requires simulations with a higher particle number.

The three different baryon models from \cite{Schneider2018} have the biggest impact on the $S_8$ constraints.
They shift the predicted $S_8$ values by approximately one standard deviation.
There is only a small difference between the three baryon models, A, B, and C, corresponding to different values of the hydrostatic mass bias.
In \cite{Schneider2018}, they found that model A had the least impact on the power spectrum, and models B and C give larger deviation from the dark matter only case.
Our results show a similar trend: the $S_8$ deviation is smallest for model A, and larger for models B and C.
In the future analyses using CNNs, it may become necessary to include realistic baryon models in the simulations.

Changing the Hubble parameter of the mock observation does not introduce a significant shift to the $S_8$ constraints. We, therefore, conclude that our inference pipeline should not be affected by the discrepancy of the measured $H_0$ values from the KiDS-450 analysis \citetalias{Hildebrandt2016} and Planck 2018 \cite{Planck2018_res}.

\section{Results \label{sec:results}}

\begin{table}
\begin{tabular}{ccc}
\hline
Parameter & \hspace{25pt}CNN\hspace{25pt} & Power spectrum \\
\hline \vspace*{-9pt}\\
\vspace*{5pt}
$S_8$ & $0.777^{+0.038}_{-0.036}$ & $0.774^{+0.060}_{-0.046}$ \\
\vspace*{5pt}
$A_\mathrm{IA}$ & $1.399^{+0.779}_{-0.724}$ & $1.619^{+1.259}_{-0.697}$ \\
\vspace*{5pt}
$\sigma_8$ & $0.724^{+0.061}_{-0.100}$ & $0.759^{+0.068}_{-0.165}$ \\
\vspace*{5pt}
$\Omega_\mathrm{m}$ & $0.357^{+0.074}_{-0.062}$ & $0.334^{+0.138}_{-0.049}$ \\
\hline
\end{tabular}
\caption{Constraints (68\% CL) of the cosmological parameters and the intrinsic alignment amplitude obtained with our CNN and power spectrum analysis.
}
\label{tab:cons}
\end{table}
\begin{figure*}
\includegraphics[width=1.0\textwidth]{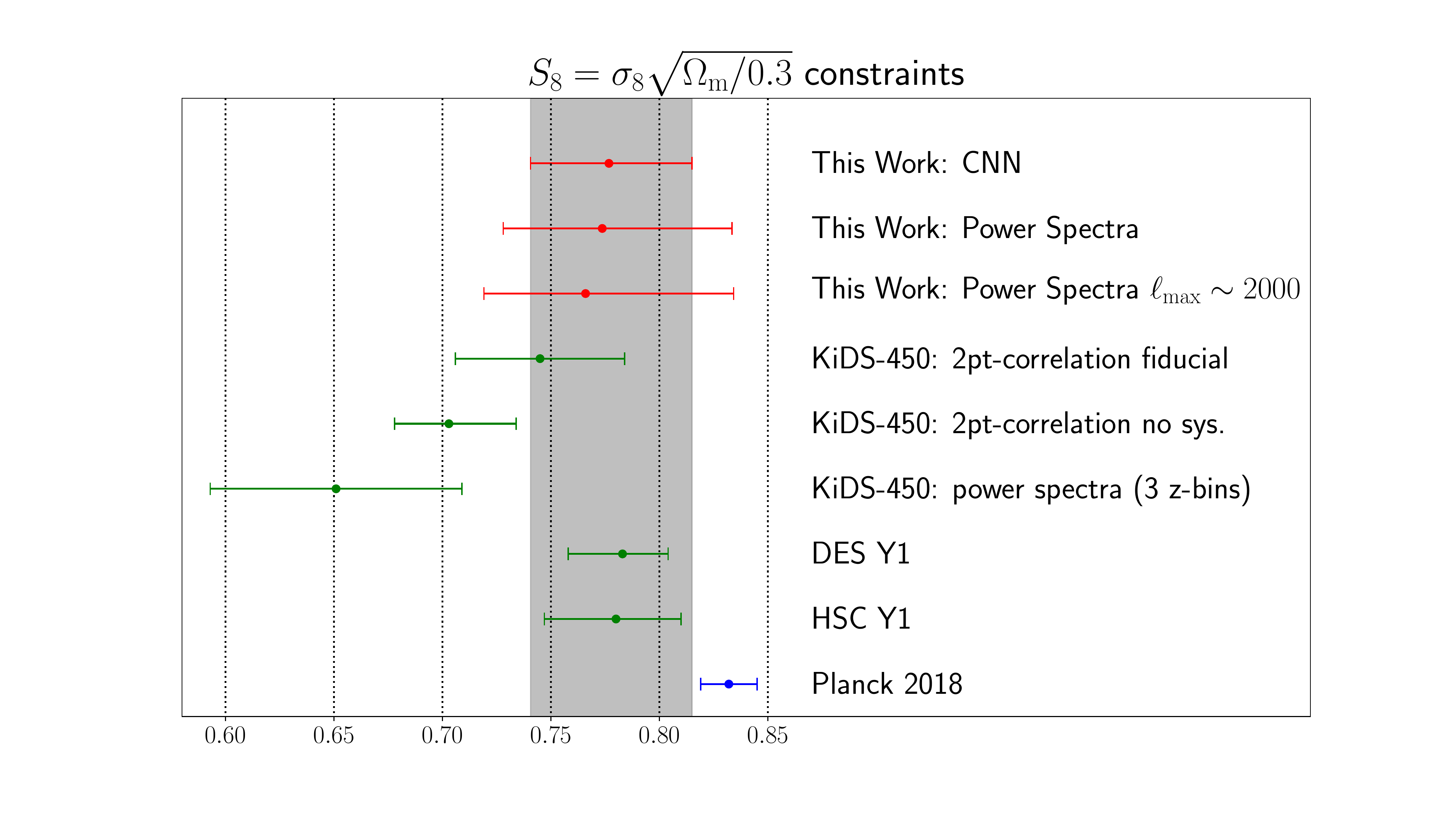}
\caption{The $S_8 = \sigma_8\sqrt{\Omega_\mathrm{m}/0.3}$ constraints of our analysis compared to the KiDS-450 two-point-correlation function analysis \citetalias{Hildebrandt2016}, the KiDS-450 power spectrum analysis \cite{KIDSpower}, the first year results of the Dark Energy Survey \cite{DESY1res}, the Subaru Hyper Suprime-Cam first year results \cite{HSCresults} and Planck 2018 \cite{Planck2018_res}. \label{fig:S8_cons}}
\end{figure*}

The constraints on the cosmological parameters and the intrinsic alignment amplitude are listed in table \ref{tab:cons}. The cosmological constraints in the $\Omega_\mathrm{m}-\sigma_8$ plane obtained with our networks and the tomographic power spectrum analysis are shown in the left panel of figure \ref{fig:CNN_vs_other} and a comparison of our constraints to previous work is shown in the right panel of figure \ref{fig:CNN_vs_other}.
We constrain the degeneracy parameter to
${S_8=0.777^{+0.038}_{-0.036}}$(68\% CL)
with our networks and to
${S_8=0.774^{+0.060}_{-0.046}}$
with our power spectrum analysis, which is also shown in figure \ref{fig:S8_cons}.

The comparison of our power spectrum analysis to the networks clearly shows that the CNN is able to extract more information than the power spectrum.
The area of the 95\% confidence contours is $\sim 50\%$ smaller for the network analysis, while the constraints on the $S_8$ parameter improve by $\sim30\%$.
These gains are consistent with the results from our mock observations and our previous work \citep{Fluri2018}.
The CNN is able to better constrain the $\sigma_8-\Omega_\mathrm{m}$ plane than the PS in the degeneracy direction, which is expected from a method exploiting the non-Gaussian features in the data.

Further, we performed another robustness test to check the impact of including the small scales on our constraints (shown in figure \ref{fig:S8_cons}).
These scales can be affected by simulation resolution effects, and as we do not use a hard cut on $\ell$, but rather use a large $\ell$ range from previously smoothed maps.
We performed another power spectrum analysis with only 6 bins, instead of 7 as in out fiducial analysis, by removing the highest $\ell$-bin.
The resulting constraints were consistent with our fiducial analysis and only slightly broader, which indicates that the constrains are not driven by the small scales.
This is expected since the used pixel size, smoothing scale and the measurement noise should have already removed a lot of information from these scales.

In the comparison with the other constraints it is important to highlight the differences in the analysis of the KiDS-450 fiducial results from the two-point-correlation \citetalias{Hildebrandt2016}, the KiDS-450 power spectrum analysis \cite{KIDSpower}, and the constraints from our networks. The KiDS-450 power spectrum analysis used only three redshift bins and a multipole range of $76 \leq \ell \leq 1310$, while the KiDS-450 two-point-correlation function analysis used four redshift bins and angular separations from 0.5$'$ to 72$'$. The two-point-correlation function had, therefore, access to much smaller scales, which explains the size difference of the corresponding contours. The relevant scales for our analysis are our pixel scale of 2.34 arcmin and the applied Gaussian smoothing kernel with $\sigma_s = 2.34$ arcmin, which lies between the smallest scales used in the two-point correlation function and the power spectrum analysis. Comparing the size of the 95\% confidence contours from our tomographic power spectrum analysis to the other constraints shows the same trend: our constraints are narrower than the ones from the PS analysis \citep{KIDSpower}, but larger than \citetalias{Hildebrandt2016}.
In our work we use the same effective redshift distributions and similar multiplicative shear bias estimates, as well as their corresponding uncertainties.
Our shear catalog was similar, with the exception of $\sim$2 million galaxies that lied outside the set of 20 flat patches we used.

It is important to note that both the KiDS-450 two-point-correlation function analysis \citetalias{Hildebrandt2016} and power spectrum analysis \cite{KIDSpower} used more cosmological and nuisance parameters, including baryon density $\Omega_b$, spectral index $n_s$, Hubble parameter $H_0$, and baryon feedback amplitude $B$.
Marginalizing over these parameters broadens the contours compared to the analysis that includes only $\sigma_8$, $\Omega_\mathrm{m}$, and $A_{\rm{IA}}$ parameters; marginalization of baryon feedback alone leads to 20\% degradation of the $S_8$ constraint.
Our constraints are mostly consistent with the KiDS-450 fiducial analysis \citetalias{Hildebrandt2016}, lying slightly above their fiducial result.
Our obtained $S_8$ constraints are higher than the KiDS-450 power spectrum analysis \cite{KIDSpower}.
A possible reason for these apparent differences is the measured intrinsic alignment amplitude $A_\mathrm{IA}$.
The intrinsic alignment amplitude $A_\mathrm{IA}$ and the degeneracy parameter $S_8$ share a positive correlation, as can be seen in figure \ref{fig:IA_cons} in appendix~\ref{ap:IA_amp}.
With our CNN analysis, we measure the intrinsic alignment amplitude
${A_\mathrm{IA}=1.399^{+0.779}_{-0.724}}$,
which is slightly higher than the the KiDS-450 two-point-correlation function analysis \citetalias{Hildebrandt2016}
${A_\mathrm{IA} = 1.10^{+0.68}_{-0.54}}$,
and much higher than the from the KiDS-450 power spectrum analysis of
${A_\mathrm{IA}= -1.72^{+1.49}_{-1.25}}$.
We therefore suspect, this difference to be partial responsible for the differences in the constraints on $S_8$.
The difference in the measurement of $A_\mathrm{IA}$ between our PS analysis and the one in \cite{KIDSpower} could also potentially come from the different redshift bin configurations.
We also notice a small shift towards higher $\Omega_\mathrm{m}$, consistent with the uncertainties.

There are multiple possible explanations for the remaining small differences. Baryonic feedback could potentially shift our $S_8$ constraints by as much as $1\sigma$. The additional cosmological parameters considered in the KiDS-450 fiducial analysis \citetalias{Hildebrandt2016} could potentially broaden and shift the constraints as well. And lastly, the chosen approach of projecting the shear catalog onto 20 independent patches using only a subset of the available galaxies could also have an impact.

\section{Conclusions}

In this work we present the first cosmological constraints from weak lensing maps obtained using convolutional neural networks and the publicly available KiDS-450 dataset. To train the CNN we used state-of-the-art cosmological simulations generated with the \Pkdgrav\ code.
We use the effective redshift distributions and multiplicative shear bias estimates to marginalize over these systematic effects.
Furthermore, we implement the commonly used intrinsic alignment model by \cite{Hirata2004, Bridle2007, Joachimi2011} on map level, enabling us to also constrain the intrinsic alignment amplitude $A_\mathrm{IA}$.

We test the impact of other possible systematic effects that can affect the analysis. We consider baryon effects, simulation configurations and engines, projection method (Born approximation vs ray-tracing), and a different value for the Hubble parameter.
We find a very small impact of these effects, except for baryon feedback, which can result in as much as $1\sigma$ shifts of the constraints.
As implementations of baryon effects in N-body simulations is computationally challenging, we are not able to marginalize over the uncertainty on Baryons in this work, but this can potentially be implemented in the future.

We trained three different residual networks with 5, 10 and 25 residual blocks (see figure \ref{fig:resblock}) and found the best constraints by combining their predictions.
Following a blinding strategy, the CNN analysis gives ${S_8=0.777^{+0.038}_{-0.036}}$ and is consistent with the constraints generated from a power spectrum analysis performed on the same data-set.
The CNNs shrink the area of the 95\% confidence intervals on the $\sigma_8~-~\Omega_\mathrm{m}$ plane by $\sim 50\%$ and constraints on the $S_8$ parameter by $\sim 30\%$, compared to the power spectrum performed on the same maps.
The improvement comes mostly from improved capacity of the CNN to break the $\sigma_8-\Omega_\mathrm{m}$ degeneracy by extracting more non-Gaussian information from the lensing maps.
Our analysis is broadly consistent with the original KiDS-450 fiducial analysis \citetalias{Hildebrandt2016}, giving slightly higher value of $S_8$, while constraining $A_{\rm{IA}}$ to be also slightly higher, on the level of $A_\mathrm{IA} = 1.399^{+0.779}_{-0.724}$.
Higher measured intrinsic alignment parameter can also provide explanation for the difference between our CNN and PS results compared to the power spectrum analysis of \citep{KIDSpower}.

We show that generating cosmological constraints from pure forward-modeling simulations is computationally feasible for our chosen parameter set and consistent with constraints generated from theoretical predictions.
Future work should include more nuisance and cosmological parameters. To do this, it is important to further improve the efficiency of the inference pipeline and to generate more N-body simulations. Recent work on cosmological emulators \cite{EUCLIDEmu} and the release of new N-body simulations \cite{HACCData} suggests that the simulation-driven inference of increasing number of cosmological parameters is becoming feasible.

Importantly, the simulation-level implementations of realistic baryon feedback models will most likely prove to be crucial for future analyses with deep learning.
Approximate methods mimicking the effects of baryons on the density field or full hydrodynamical simulations have the potential to address this issue \citep{baryonpainting,Schneider2018}.

Further examinations of the scalability of the inference pipeline are also important to apply it to larger data-sets such as the Dark Energy Survey. The training of larger networks becomes possible through parallelized training strategies and the geometry of large survey can be taken into account with network architectures such as graph-based networks \cite{Perraudin2018}. Finally, it would be interesting to compare and combine the constraints generated by the CNN with the power spectrum and higher order statistics.

\begin{acknowledgments}
This work was supported by the Swiss Data Science Centre (SDSC), project \textit{sd01 - DLOC:  Deep Learning for Observational Cosmology}, and grant number 200021\_169130 and PZ00P2\_161363 from the Swiss National Science Foundation. We thank the KiDS collaboration for publishing their data products with a very good quality of documentation.
\end{acknowledgments}

\appendix


\section{Multiple CNN vs single CNN}
\label{sec:multi_vs_single_cnn}

\begin{figure}
\includegraphics[width=0.5\textwidth]{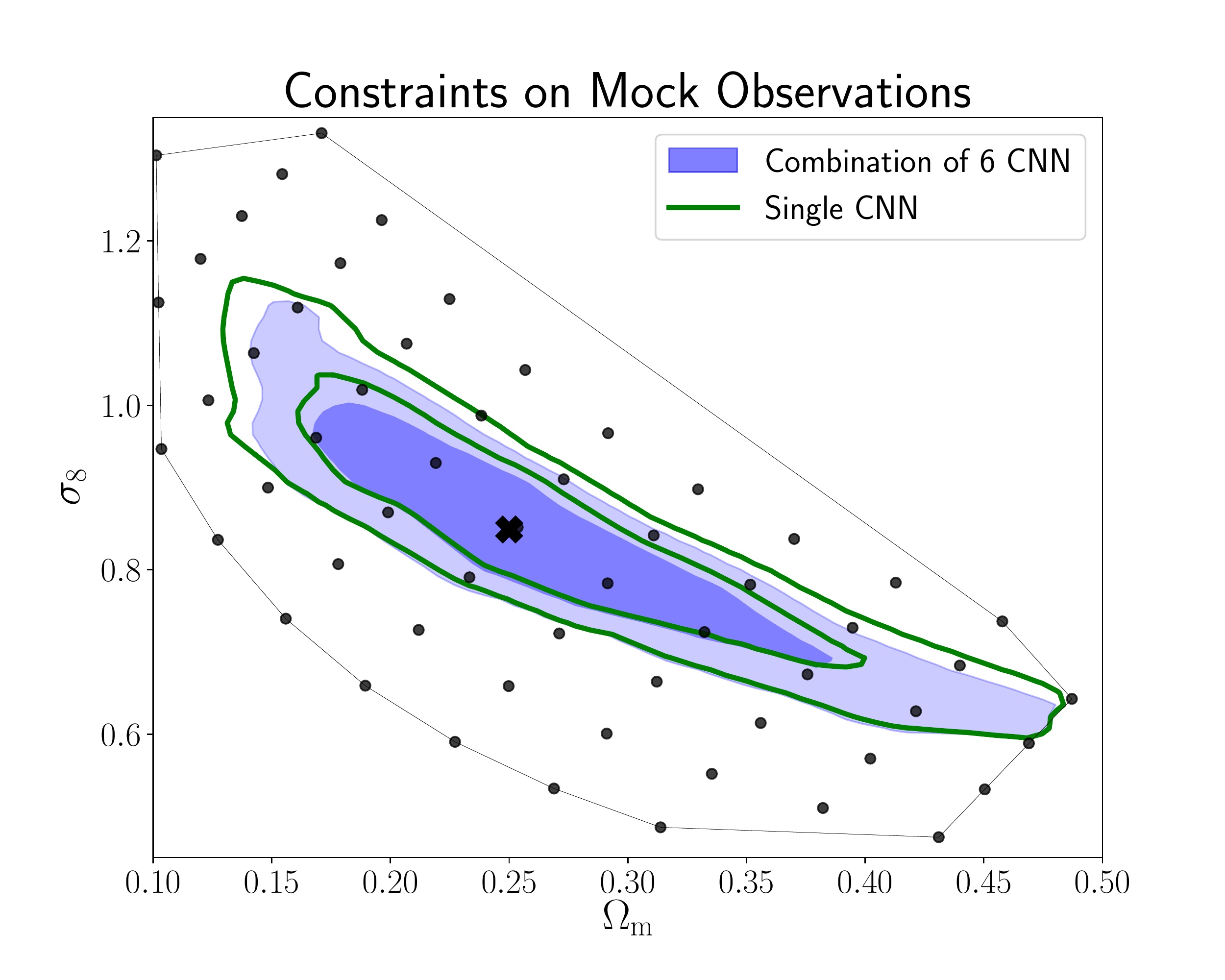}
\caption{The 68\% and 95\% confidence contours of our fiducial mock observation obtained by combining the predictions of all our six considered networks (blue) and a single network (green). The convex hull of our simulated cosmologies (prior) is drawn in black. The black cross in the middle shows the cosmological parameters used to generate the mock observations. \label{fig:fiducial_cons}}
\end{figure}

As described in section~\ref{sec:combining_networks}, a combination of separate CNNs results in improved constraints.
The outputs of these networks was then concatenated and used in further likelihood analysis as a large data vector.
In figure~\ref{fig:fiducial_cons} we show the comparison between constraints from a single network and combination of multiple networks.

\section{Train and Validation Loss \label{ap:loss}}

\begin{figure*}
\includegraphics[width=1.0\textwidth]{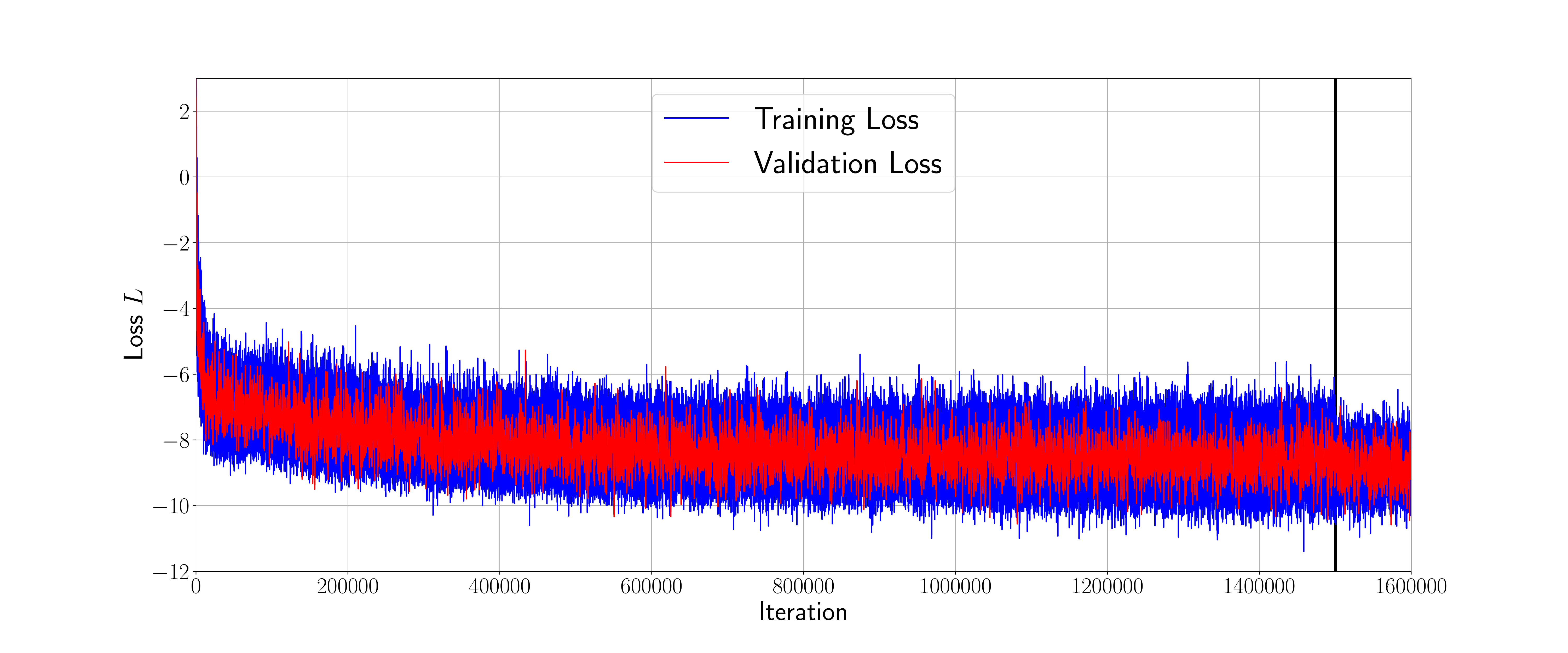}
\caption{The train and validation loss of the network architecture with 15 residual blocks. The black solid line indicates the switch from asynchronous training to synchronous training. \label{fig:loss}}
\end{figure*}

An example of the training and validation loss of the network built out of 15 residual blocks is shown in figure \ref{fig:loss}. The two losses are almost identical over the whole training, which indicates that the network do not overfit to the training-set. The small difference in the loss for the asynchronous and synchronous training does not significantly affect the resulting cosmological constraints.

\section{Evaluation of the Test Set \label{ap:evals}}

\begin{figure*}
\includegraphics[width=1.0\textwidth]{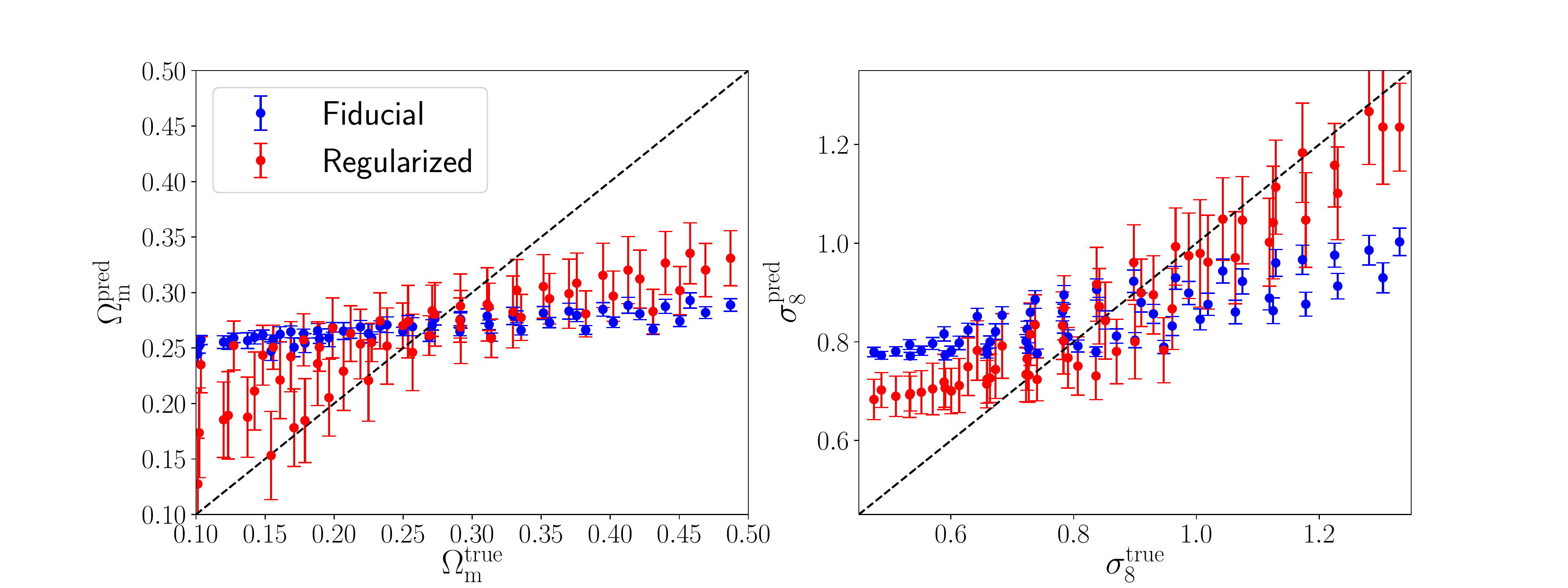}
\caption{Evaluations of the test set from a network with 15 residual blocks. The blue points correspond to our fiducial training with the loss described in equation \ref{eq:likelihoodloss}. For the red points we add the loss of equation \ref{eq:avgloss} with $\lambda = 1000$. \label{fig:preds}}
\end{figure*}

The predictions on the test set for the network with 15 residual blocks are shown in figure \ref{fig:preds}. These predictions are heavily biased towards the center of the grid. However, this bias does not affect the resulting cosmological constraints, since we are using the predictions as summary statistics. The bias can be reduced by introducing a square loss over the average prediction of the 20 projected patches
\begin{equation}
L_2 = \lambda\left\lVert\left(\frac{1}{20}\sum_{i=1}^{20}\theta_i^p\right) - \theta^t\right\lVert^2, \label{eq:avgloss}
\end{equation}
where $\lambda$ is a constant chosen such that the loss is of the same order of magnitude as the negative log-likelihood loss of equation \ref{eq:likelihoodloss}. The predictions of a network with 15 residual blocks, which is trained with the standard likelihood loss and the square loss with $\lambda = 1000$, are shown in figure \ref{fig:preds}. It can be seen that, while the bias is reduced, the variance of the predictions increases. We do not choose this combination loss in our fiducial analysis, since the resulting cosmological constraints are slightly worse.

\section{Intrinsic Alignment Amplitude \label{ap:IA_amp}}

\begin{figure}
\includegraphics[width=0.5\textwidth]{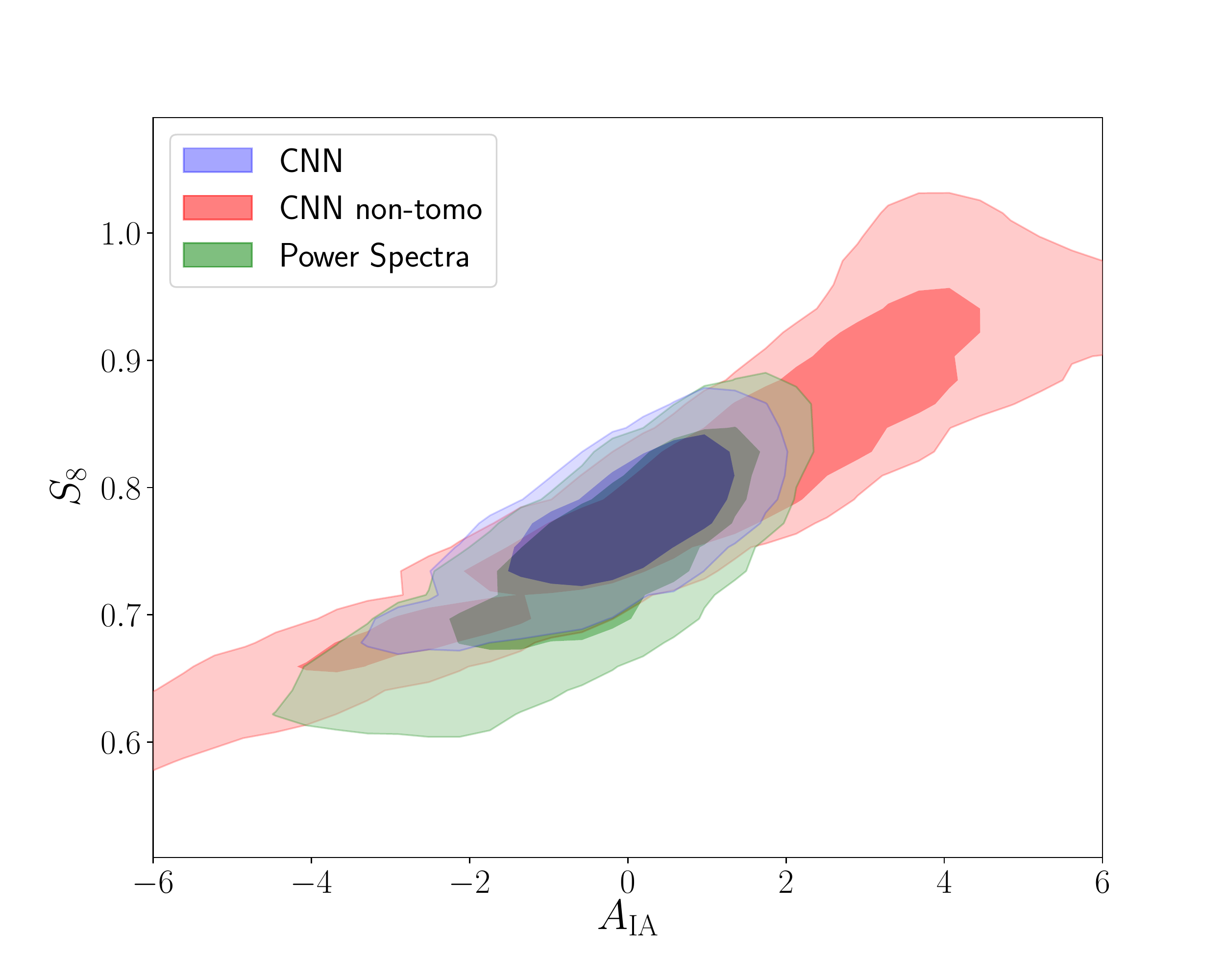}
\caption{The constraints on the intrinsic alignment amplitude $A_\mathrm{IA}$ and the degeneracy parameter $S_8$ for our fiducial analysis, a non-tomographic network analysis and our power spectrum analysis using our fiducial mock observation. \label{fig:IA_cons} }
\end{figure}

In figure \ref{fig:IA_cons} we show the constraints on the intrinsic alignment amplitude $A_\mathrm{IA}$ and the degeneracy parameter $S_8$ for different settings using our fiducial mock observation. One can clearly see the large degeneracy between the cosmological parameters and the intrinsic alignment amplitude in the non-tomographic case. Performing a tomographic analysis helps to break this degeneracy for the CNN and the power spectrum analysis.
Using our fiducial mock observation with $A_\mathrm{IA} = 0$ we obtain the constraints $A_\mathrm{IA} = -0.1 \pm 1.0$, $A_\mathrm{IA} = -0.4 \pm 1.4$ for our power spectrum analysis and $A_\mathrm{IA} = -0.7 \pm 2.6$ for our non-tomographic analysis with a single network.

\section{Power spectrum: Cross-correlations and Covariance Matrix \label{ap:specs}}

\begin{figure*}
\includegraphics[width=1.0\textwidth]{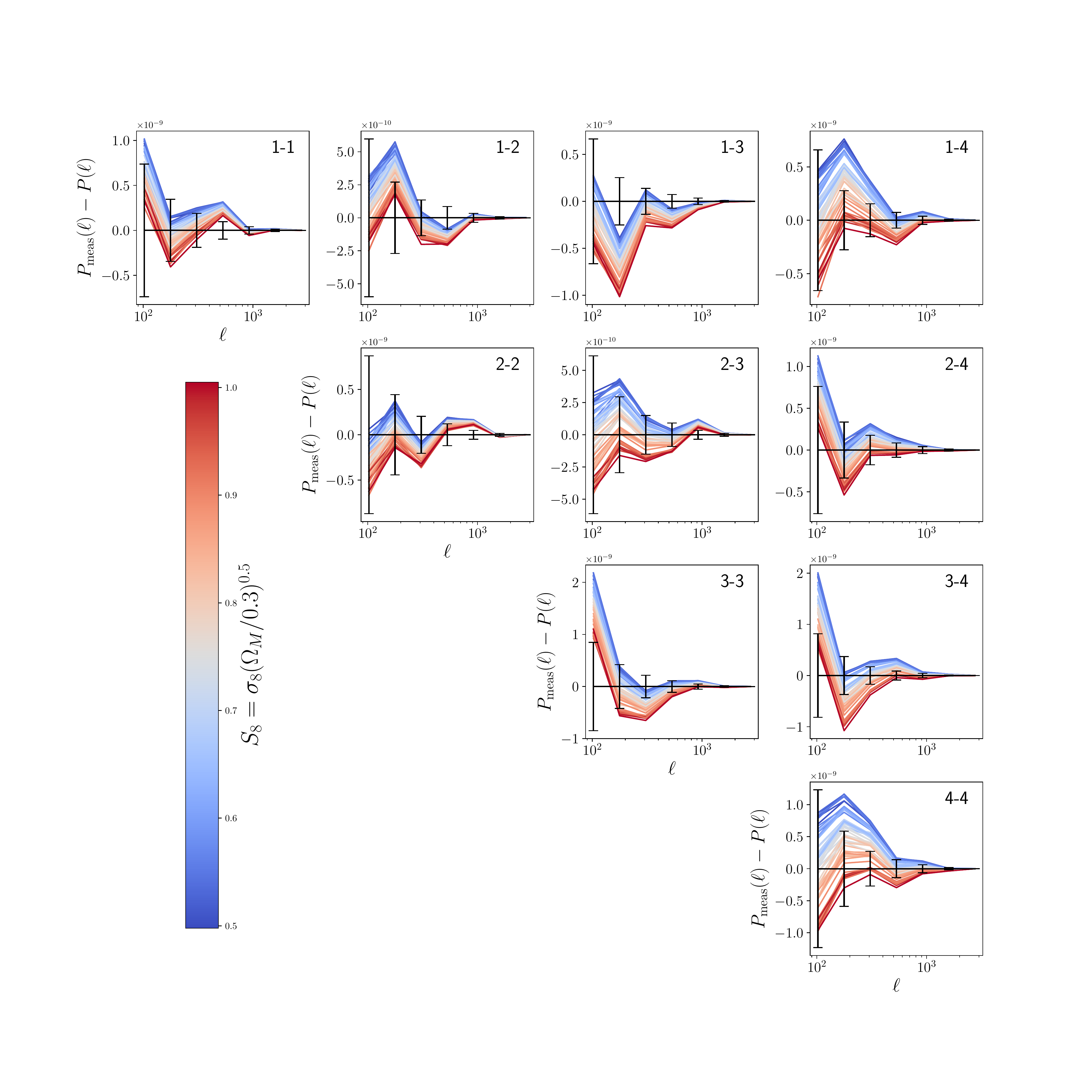}
\caption{Difference of the measured power spectrum of the KiDS-450 data-set and the average auto- and cross-correlations of the all simulated cosmologies from the simulated surveys used for the inference with intrinsic alignment amplitude $A_\mathrm{IA} = 1.5$. The upper left corner shows the auto-correlation of the first redshift bin and the lower right corner from the fourth redshift bin. The spectrum includes the applied noise, mask and smoothing. The errorbars are obtained from the diagonal entries of the used covariance matrix (see fig. \ref{fig:powcov}) \label{fig:specs}}
\end{figure*}

\begin{figure}
\includegraphics[width=0.5\textwidth]{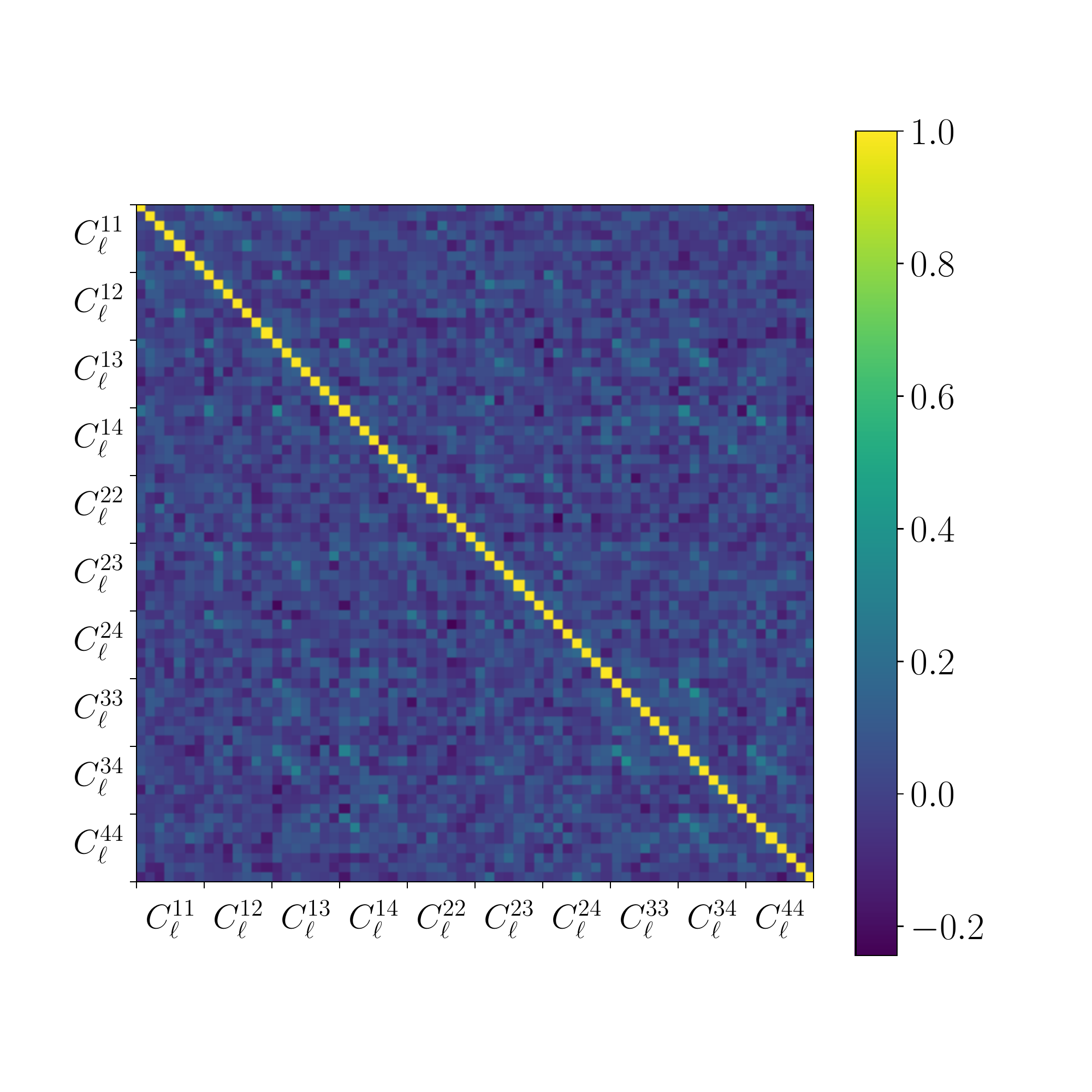}
\caption{Correlation matrix obtained from the covariance matrix used for the power spectrum analysis. \label{fig:powcov}}
\end{figure}

We show a comparison of our measured power spectrum from the projected KiDS-450 data averaged over the 20 patches with the power spectrum from our simulations used for the inference in figure \ref{fig:specs}. Since we did average over all patches with different masks and applied Gaussian smoothing, it is difficult to disentangle the noise and the signal. The shown errorbars are obtained from the diagonal of the covariance matrix used for the inference. The correlation matrix (normalized covariance matrix) of the simulated surveys is shown in figure \ref{fig:powcov}. Again, the matrix is difficult the interpret, as it contains noise, smoothing and the average over all 20 projected patches.

\newpage

\bibliography{library}

\end{document}